\documentclass[fleqn,usenatbib]{mnras}

\usepackage{newtxtext,newtxmath}

\usepackage[T1]{fontenc}

\DeclareRobustCommand{\VAN}[3]{#2}
\let\VANthebibliography\thebibliography
\def\thebibliography{\DeclareRobustCommand{\VAN}[3]{##3}\VANthebibliography}

\usepackage{graphicx}	
\usepackage{amsmath}	
\usepackage{fdsymbol}

\title[Cosmological simulations of SIDM]{Cosmological simulations with rare and frequent dark matter self-interactions}

\author[M. S. Fischer et al.]{Moritz S. Fischer,$^{1}$\thanks{E-mail: moritz.fischer@uni-hamburg.de (UHH)}
Marcus Br\"{u}ggen,$^{1}$
Kai Schmidt-Hoberg,$^{2}$
Klaus Dolag,$^{3,4}$
\newauthor{Felix Kahlhoefer,$^{5,6}$
Antonio Ragagnin,$^{7,8,9}$
Andrew Robertson$^{10}$}
\\
$^{1}$Hamburger Sternwarte, Universit\"at Hamburg, Gojenbergsweg 112, D-21029 Hamburg, Germany\\
$^{2}$Deutsches Elektronen-Synchrotron DESY, Notkestr.~85, 22607 Hamburg, Germany\\
$^{3}$Universit\"ats-Sternwarte, Fakult\"at f\"ur Physik, Ludwig-Maximilians-Universit\"at M\"unchen, Scheinerstr. 1, D-81679 M\"unchen, Germany\\
$^{4}$Max-Planck-Institut f\"ur Astrophysik, Karl-Schwarzschild-Str. 1, D-85748 Garching, Germany\\
$^{5}$Institute for Theoretical Particle Physics and Cosmology (TTK), RWTH Aachen University, D-52056 Aachen, Germany\\
$^{6}$Institute for Theoretical Particle Physics (TTP), Karlsruhe Institute of Technology (KIT),
D-76128 Karlsruhe, Germany\\
$^{7}$Dipartimento di Fisica e Astronomia "Augusto Righi", Alma Mater Studiorum Università di Bologna, via Gobetti 93/2, I-40129 Bologna, Italy\\ 
$^{8}$INAF-Osservatorio Astronomico di Trieste, via G. B. Tiepolo 11, I-34143 Trieste, Italy\\
$^{9}$IFPU -- Institute for Fundamental Physics of the Universe, Via Beirut 2, I-34014 Trieste, Italy\\
$^{10}$Jet Propulsion Laboratory, California Institute of Technology, 4800 Oak Grove Drive, Pasadena, CA 91109, USA
}

\date{Accepted XXX. Received YYY; in original form ZZZ}

\pubyear{2022}

\begin{document}
\label{firstpage}
\pagerange{\pageref{firstpage}--\pageref{lastpage}}
\maketitle

\begin{abstract}
    Dark matter (DM) with self-interactions is a promising solution for the small-scale problems of the standard cosmological model.
    Here we perform the first cosmological simulation of \emph{frequent} DM self-interactions, corresponding to small-angle DM scatterings.
    The focus of our analysis lies in finding and understanding differences to the traditionally assumed \emph{rare} DM (large-angle) self-scatterings.
    For this purpose, we compute the distribution of DM densities, the matter power spectrum, the two-point correlation function, and the halo and subhalo mass functions.
    Furthermore, we investigate the density profiles of the DM haloes and their shapes.
    We find that overall large-angle and small-angle scatterings behave fairly similarly with a few exceptions.
    In particular, the number of satellites is considerably suppressed for frequent compared to rare self-interactions with the same cross-section.
    Overall, we observe that while differences between the two cases may be difficult to establish using a single measure, the degeneracy may be broken through a combination of multiple ones.
    For instance, the combination of satellite counts with halo density or shape profiles could allow discriminating between rare and frequent self-interactions.
    As a by-product of our analysis, we provide -- for the first time -- upper limits on the cross-section for frequent self-interactions.
\end{abstract}




\section{Introduction} \label{sec:introduction}
Although many efforts have been made to uncover the nature of dark matter (DM), it remains largely unknown even after several decades of research.
To narrow down the large number of models that contain potential DM candidates, a huge variety of experiments, based on direct and indirect detection, are being carried out.
Moreover, forthcoming astronomical surveys with upcoming telescopes such as \textsc{Euclid}\footnote{\textsc{Euclid}: \url{https://www.euclid-ec.org/}} \citep{Euclid_Collaboration_2020}, Rubin Observatory\footnote{Rubin Observatory: \url{https://www.lsst.org/}} \citep{Zhan_2018} and Roman\footnote{Nancy Grace Roman Space Telescope: \url{https://www.jpl.nasa.gov/missions/the-nancy-grace-roman-space-telescope}} \citep{Spergel_2015} promise to tighten constraints on cosmological models and to discriminate between models of DM beyond the cold collisionless DM of the standard cosmological model (Lambda cold dark matter, $\Lambda$CDM).
Among those are warm DM \citep{Dodelson_1994} and fuzzy DM \citep{Hu_2000}.

In this paper, we focus on a particular class of DM models called self-interacting DM (SIDM). It was first proposed by \cite{Spergel_2000} in order to resolve tensions between cosmological $N$-body simulations and observations. These tensions are known as the small-scale crisis of $\Lambda$CDM \citep[for a review, see ][]{Bullock_2017}.
SIDM has been studied in a number of papers and seems to be promising to solve or at least mitigate several small-scale issues.
For a review on DM with self-interactions see \cite{Tulin_2018}.
In the limit of a vanishing cross-section, SIDM becomes identical to the collisionless DM of $\Lambda$CDM.
But given a large enough cross-section, it alters the DM distribution on small scales and may resolve issues such as the core--cusp problem.
Self-interactions can transfer heat into the centres of DM haloes and thus create density cores, in contrast to the cusps of collisionless cold dark matter (CDM). 
This has been shown, for example, in \cite{Dave_2001}. Moreover, SIDM can create diverse rotation curves \citep{Creasey_2017, Kamada_2017, Robertson_2018} and may be able to solve the too-big-to-fail problem \citep{Zavala_2013, Elbert_2015, Kaplinghat_2019a}.

There exist a number of SIDM models with a set of free parameters, such as the total cross-section, the angular and velocity dependence, and the nature of the scattering (elastic or inelastic).
Most studies have assumed models that are isotropic, elastic, and velocity-independent.
\cite{Burkert_2000} performed the first simulations with a Monte-Carlo scheme where the numerical particles were treated analogously to physical DM particles.
Since then many variants of SIDM have been studied, such as inelastic scattering \citep[e.g.][]{Essig_2019, Huo_2019, Shen_2021} including multistate scattering \citep{Schutz_2015, Vogelsberger_2019, Chua_2020} or even
multicomponent DM \citep{Todoroki_2018, Vogelsberger_2019}.
Also anisotropic cross-sections have been investigated \citep{Robertson_2017b, Banerjee_2020, Nadler_2020}.

However, if the self-interaction cross-section is strongly anisotropic, particles scatter by tiny angles. This implies a much lower momentum and energy transfer per scattering event compared to an isotropic cross-section. Hence, small-angle scattering must be more frequent to have a similar effect on the DM distribution. Models having different angular dependences might be compared by using the momentum-transfer cross-section. By frequent self-interacting DM (fSIDM), we refer to a limit where the scattering angles become infinitesimal small, while the momentum-transfer cross-section stays constant. In contrast, we refer to rare self-interacting DM (rSIDM) for less anisotropic differential cross-sections.

Frequent self-interactions have gained popularity in the context of galaxy cluster mergers because they can explain larger DM--galaxy offsets than rSIDM \citep{Kahlhoefer_2014, Fischer_2021a, Fischer_2021b}.
However, numerical schemes that treat numerical particles like physical ones are not capable of simulating fSIDM.
Only recently, a general solution to this problem has been found.
\cite{Fischer_2021a} developed a new scheme that allows one to model frequent scattering within $N$-body simulations.

However, previous fSIDM studies \citep{Kahlhoefer_2014, Kahlhoefer_2015, Kummer_2018, Kummer_2019, Fischer_2021a, Fischer_2021b} only considered idealized cases that never took the full cosmological context into account.
In contrast, for rSIDM, there are a number of recently published simulations of cosmological boxes \citep{Peter_2013, Rocha_2013, Vogelsberger_2016, Robertson_2019, Robertson_2020, Banerjee_2020, Stafford_2020, Stafford_2021, Harvey_2021, Ebisu_2022} or using zoom-in simulations \citep{Vogelsberger_2012, Vogelsberger_2016, Vogelsberger_2019, Zavala_2013, Zavala_2019, Vogelsberger_2014, Fry_2015, Robertson_2018, Despali_2019, Robles_2019, Nadler_2020, Vega-Ferrero_2020, Bondarenko_2021, Sameie_2021, Shen_2021, Shen_2022,  Bhattacharyya_2022, Silverman_2022, Sirks_2022}.
These simulations have been used to study the phenomenology of SIDM models on various mass scales, such as dwarf galaxies, Milky Way (MW)-like galaxies, and galaxy clusters.
Several properties of the DM haloes such as their density profile and their shape have been measured and predictions for observations, such as gravitational lensing, have been made.
This enabled constraints to be put on the total cross-section of rSIDM models, while fSIDM models have remained poorly constrained.
In this study, we investigate fSIDM, for the first time using a cosmological simulation.

This paper aims to study the effects of SIDM on large scales to understand the differences between fSIDM and isotropic rSIDM. In this first effort, we assume the self-interactions to be velocity-independent and elastic.
We conduct DM-only cosmological $N$-body simulations using a full box as well as zoom-in simulations and study various properties of the DM distribution.

In Section~\ref{sec:numerical_setup} we briefly describe our numerical methods and present the setup for our cosmological simulations.
The results are presented in Section~\ref{sec:results}. For example, we show the matter power spectrum, the halo mass function as well as density and shape profiles of DM haloes.
A discussion of our results, their limitations, implications and further perspectives follows in Section~\ref{sec:disscussion}.
Finally, we summarize and conclude in Section~\ref{sec:conlusion}.
Additional details and plots are provided in the appendices.

\section{Numerical Setup} \label{sec:numerical_setup}

In this section, we describe the numerical setup for this study.
We give details of the code and algorithms that we have used as well as our simulations and their initial conditions. 

In this paper, we used the cosmological $N$-body code \textsc{gadget-3}, and the predecessor \textsc{gadget-2} is described in \cite{Springel_2005}.
The implementation of rare and frequent self-interactions has previously been described in \cite{Fischer_2021a, Fischer_2021b}.
Additionally, we implemented the comoving integration for the SIDM module to perform cosmological simulations.
A test problem that demonstrates that the comoving integration is working as expected can be found in Appendix~\ref{sec:comoving_integration_test}.
To match rare and frequent self-interactions, we use the momentum-transfer cross-section,\footnote{We implicitly assume identical particles, in this case, the definition is equivalent to the one recommended by \cite{Robertson_2017b} and \cite{Kahlhoefer:2017umn}.
}
\begin{equation} \label{eq:momentum_transfer_cross_section}
    \sigma_\mathrm{\tilde{T}} = 4\pi \int_0^1 \frac{\mathrm{d} \sigma}{\mathrm{d} \Omega_\text{cms}} (1 - \cos \theta_\text{cms}) \mathrm{d} \cos \theta_\text{cms} \, .
\end{equation}

We simulated a full cosmological box and and also performed zoom-in simulations.
All simulations are DM only and the self-interactions are always velocity-independent and elastic.
In the case of rSIDM, the differential cross-section is isotropic, while fSIDM corresponds to a very anisotropic cross-section.

The size of the self-interaction kernel for each particle is set by the distance to the 64th nearest neighbour. Finally, we employ the following cosmological parameters: $\Omega_\mathrm{M} = 0.272$, $\Omega_\mathrm{\Lambda} = 0.728$, $h=0.704$, $n_s = 0.963$, and $\sigma_8 = 0.809$ \citep[WMAP7;][]{Komatsu_2011}.

To generate the initial conditions for the full box, we use \textsc{N-GenIC} \citep{ngenic}.
The initial conditions are similar to box4 of the Magneticum simulations\footnote{Magneticum: \url{http://www.magneticum.org}} with a comoving side length of $48 \, \mathrm{Mpc} \, h^{-1}$.
We run simulations with different resolutions and refer to those using the naming convention of Magneticum (hr and uhr).
Our highest resolution run (uhr) contains $\sim 1.9 \times 10^8$ simulation particles.
More details on the full cosmological box are given in Tab.~\ref{tab:sim_props_box}.
Moreover, we performed cosmological zoom-in simulations with different resolutions of the same region.
The region is selected from a large box with a comoving side length of $1 \, \mathrm{Gpc}\,h^{-1}$.
Several publications \citep[e.g.][]{Planelles_2014, Rasia_2015} have used this box for zoom-in initial conditions and it was first described in \cite{Bonafede_2011}.
In our zoom-in region, the most massive halo has a virial mass of $\sim 8.8 \times 10^{11} \, \mathrm{M_\odot} \, h^{-1}$.
Further details can be found in Tab.~\ref{tab:sim_props_zoom}.
In addition, we provide in Appendix~\ref{sec:density_conv} a convergence test of the density profile of the most massive halo.

For the analysis, we identify DM haloes using the friends-of-friends algorithm\footnote{A description of the friends-of-friends algorithm can, for example, be found in \cite{More_2011}.} implemented along with \textsc{gadget-3}.
The built-in module \textsc{subfind} also identifies substructure within the haloes \citep{Springel_2001, Dolag_2009}.
We use halo and subhalo positions, masses and radii as provided by \textsc{subfind}.
The virial radius, $r_\mathrm{vir}$, and the virial mass, $M_\mathrm{vir}$, are measured with the spherical-overdensity approach based on the overdensity predicted by the generalized spherical top-hat collapse model \cite[e.g.][]{Eke_1996}.
Here, $r_\mathrm{vir}$ is defined as the radius at which the mean density becomes larger than the one of the top-hat collapse model and $M_\mathrm{vir}$ is the mass inside $r_\mathrm{vir}$.
Every halo contains at least one subhalo, which is the primary subhalo located at the same position as the halo (determined by the location of the most gravitationally bound particle).
The primary subhalo typically contains most of the particles that belong to the halo.

\begin{table}
    \centering
    \begin{tabular}{c|c|c|c|c}
        Name & $l_\mathrm{box}$ & $N_\mathrm{DM}$ & $m_\mathrm{DM}$ &  $\sigma_\mathrm{\Tilde{T}}/m_\chi$ \\
        & $(\mathrm{cMpc} \, h^{-1})$ &    & $(\mathrm{M_\odot} \, h^{-1})$ & $(\mathrm{cm}^2 \, \mathrm{g}^{-1})$ \\ \hline
        hr & 48 & $216^3$ & $8.28 \times 10^8$ & 0.0, 0.1, 1.0\\
        uhr & 48 & $576^3$ & $4.37 \times 10^7$ & 0.0, 0.1, 1.0
    \end{tabular}
    \caption{Properties of the full cosmological box simulations. In detail we provide the name, the side length of the comoving box ($l_\mathrm{box}$), the number of numerical DM particles ($N_\mathrm{DM}$), and the mass of the numerical DM particles ($m_\mathrm{DM}$) as well as the momentum-transfer cross-section per physical DM particle mass ($\sigma_\mathrm{\Tilde{T}}/m_\chi$).
    The non-zero cross-sections have been simulated using fSIDM and isotropic rSIDM. All simulations share the same initial conditions but with a different resolution.}
    \label{tab:sim_props_box}
\end{table}

\begin{table}
    \centering
    \begin{tabular}{c|c|c|c|c}
        name & $N_\mathrm{high\,res}$ & $m_\mathrm{DM}$ & $\sigma_\mathrm{\Tilde{T}}/m_\chi$ \\
        & & $(\mathrm{M_\odot} \, h^{-1})$ &  $(\mathrm{cm}^2 \, \mathrm{g}^{-1})$ \\
         \hline
        1x & $\sim4.51 \times 10^4$ & $8.3 \times 10^8$ & 0.0, 1.0\\
        10x & $\sim4.52 \times 10^5$ & $8.3 \times 10^7$ & 0.0, 1.0\\
        25x & $\sim1.13 \times 10^6$ & $3.3 \times 10^7$ & 0.0, 1.0\\
        250x & $\sim1.13 \times 10^7$ & $3.3 \times 10^6$ & 0.0, 1.0\\
        2500x & $\sim1.13 \times 10^8$ & $3.3 \times 10^5$ & 0.0, 1.0
    \end{tabular}
    \caption{Properties of the zoom-in simulations.
    We provide the name of the simulation, the number of particles in the highly resolved region ($N_\mathrm{high\,res}$), the mass of the high-resolution particles ($m_\mathrm{DM}$) and the cross-sections we simulated ($\sigma_\mathrm{\Tilde{T}}/m_\chi$).
    The non-zero cross-section has been simulated using fSIDM and isotropic rSIDM. All simulations share the same initial conditions but with a different resolution.}
    \label{tab:sim_props_zoom}
\end{table}

\section{Results} \label{sec:results}
In this section, we present the results of our simulations and compare the effects of DM models.
To this end, we study several statistical properties such as the matter power spectrum, the probability density function (PDF) of the DM densities, the two-point correlation function, and the halo and subhalo mass function.
We then study the impact of self-interactions on the density and circular velocity profile of DM haloes.
Furthermore, we investigate how the shapes of haloes change when self-interactions are present.
Besides, we study qualitative differences between rSIDM and fSIDM and discuss transferring constraints on the cross-section of rare scatterings to frequent self-interactions.

\subsection{Surface density}

\begin{figure*}
    \centering
    \includegraphics[width=\textwidth]{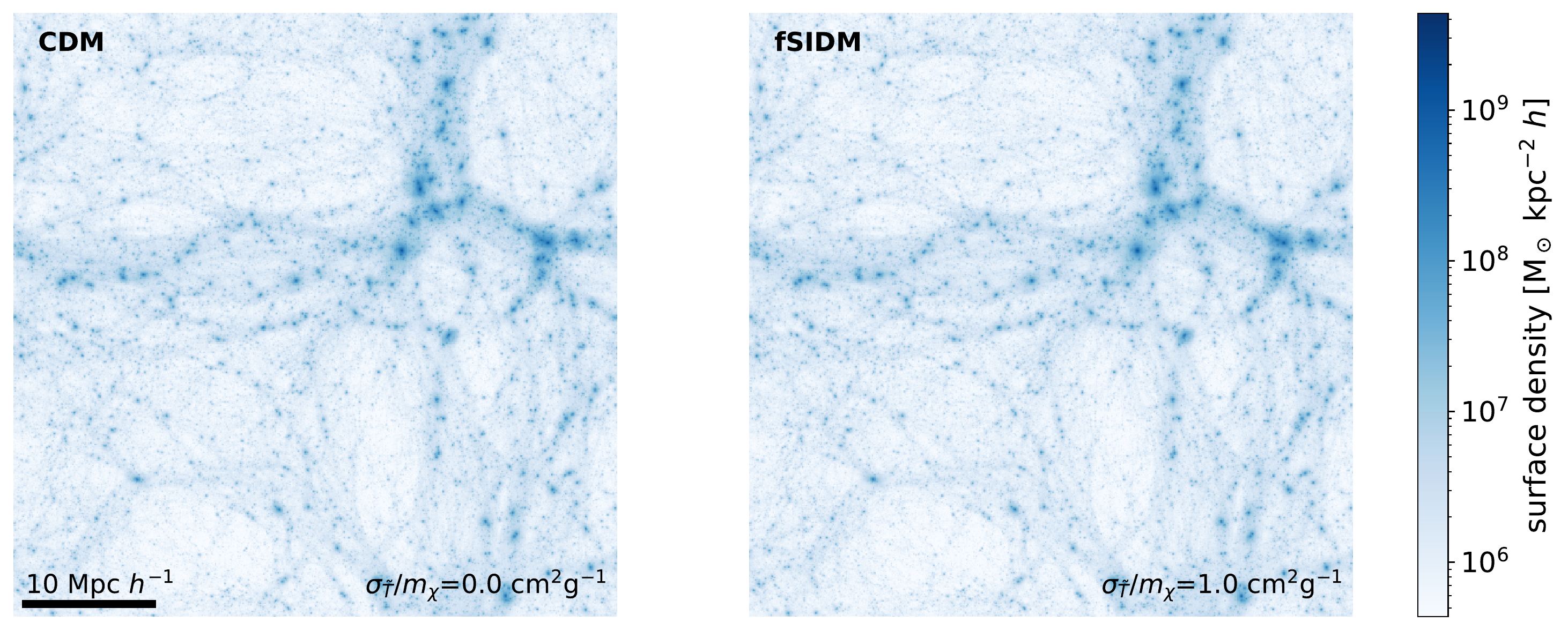}
    \caption{Surface density of the full cosmological box (uhr) at a redshift of $z=0$.
    The panel on the left-hand side shows results from the CDM simulation and the one on the right-hand side displays the fSIDM simulation with $\sigma_\mathrm{\Tilde{T}}/m_\chi = 1.0 \, \mathrm{cm}^2 \, \mathrm{g}^{-1}$.
    }
    \label{fig:density}
\end{figure*}

In Fig.~\ref{fig:density}, we show the surface density of the full cosmological box for our CDM and fSIDM ($\sigma_\mathrm{\Tilde{T}}/m_\chi = 1.0 \, \mathrm{cm}^2 \, \mathrm{g}^{-1}$) simulations.
At a cosmological redshift of $z = 0$, basically, no differences between the simulations are visible.
Hence, fSIDM seems to agree well with the collisionless DM on large scales.
Previous studies that examined the large-scale structure in rSIDM found that it looks like CDM, but differences arise on small scales \citep[e.g.][]{Rocha_2013, Stafford_2020}.
Hence, SIDM keeps the success of CDM in explaining the large-scale structure but could be capable of resolving small-scale issues.
In the next sections, we investigate quantitatively the effects of fSIDM and rSIDM.

\subsection{Matter power spectrum}

\begin{figure}
    \centering
    \includegraphics[width=\columnwidth]{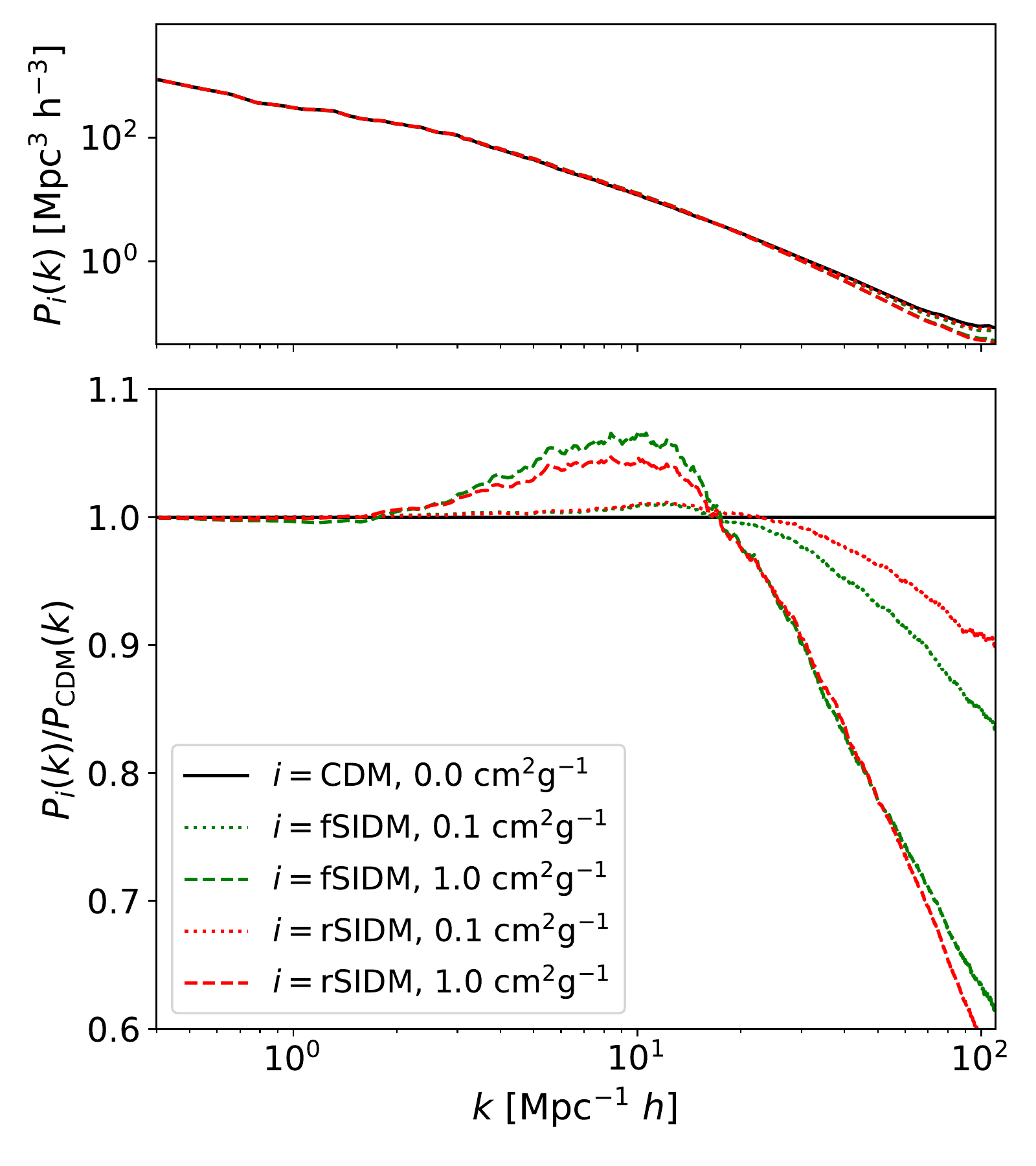}
    \caption{In the upper panel, we show the matter power spectrum for the highest resolution box (uhr), and in the lower panel we display the ratio to CDM.
    The results are for a redshift of $z=0$.
    }
    \label{fig:power_spectrum}
\end{figure}

Cosmological structure is characterized by the matter power spectrum.
\cite{Stafford_2021} computed it for several cosmologies including isotropic SIDM.
They found a suppression of small-scale structure with increasing cross-section, while on large scales SIDM behaves like CDM.

In Fig.~\ref{fig:power_spectrum}, we show the matter power spectrum of our cosmological full-box simulations and compare the various DM models.
To compute the power spectrum, we use the same code as in \cite{Grossi_2008}.
In line with \cite{Stafford_2021}, we find that the self-interactions affect the small scales (high $k$-values) only and can lead here to substantial suppression of structures.
The stronger the self-interactions, the stronger the suppression of structure formation on small scales.
However, we only find small differences between the rSIDM and fSIDM simulations.
In particular, for the larger cross-section of $\sigma_\mathrm{\Tilde{T}}/m_\chi = 1.0 \, \mathrm{cm}^2 \, \mathrm{g}^{-1}$, the effects are very similar.

\subsection{Distribution of DM densities}
\begin{figure}
    \centering
    \includegraphics[width=\columnwidth]{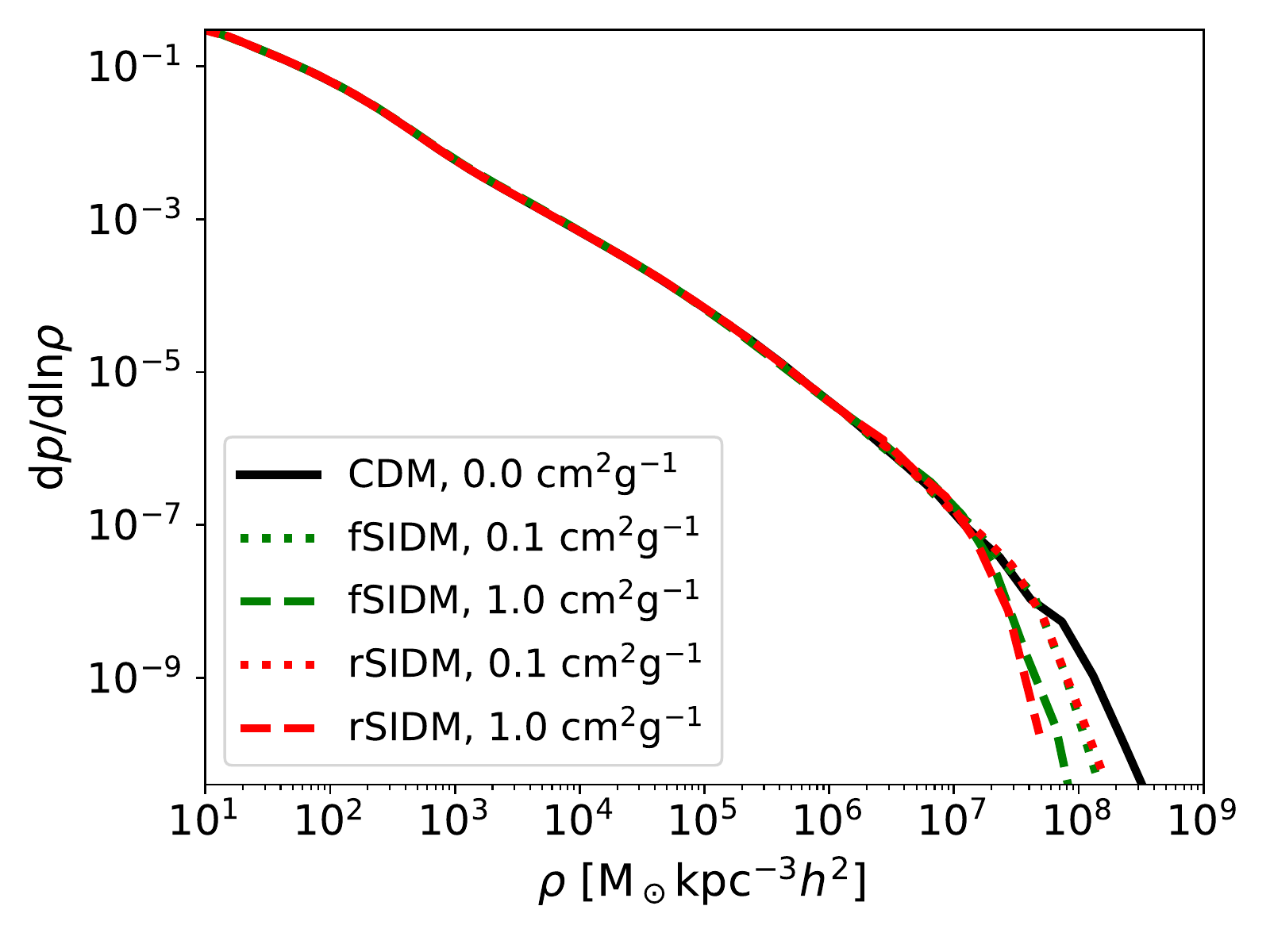}
    \caption{The probability to find a given density per logarithmic density bin is shown as a function of density for various simulations.
    The colours indicate the type of self-interaction and the line style gives the strength of self-interaction as indicated in the legend.
    The plot is for a redshift of $z=0$ and produced from the high-resolution full cosmological box simulations (uhr).
    }
    \label{fig:pdd}
\end{figure}

Here, we compute the volume-weighted probability to find a given density, i.e.\ the PDF for a random position within the cosmological volume.
First, we use an oct-tree to find neighbours of the simulation particles.
For each particle, we find the radius that contains 160 neighbouring particles (for the highest resolution run, uhr).
Using that radius we estimate a physical density and divide it by the particle mass to obtain a volume that is associated with the particle.
From all particles, we sum their associated volumes per logarithmic density bin and divide by the simulation volume and logarithmic density bin size.
The result is shown in Fig.~\ref{fig:pdd}.

We find that the various DM models differ in the high-density regime only.
Here, self-interactions suppress the highest densities compared to CDM. The suppression takes place for densities $\gtrsim 10^7 \, \mathrm{M_\odot} \, \mathrm{kpc}^{-3} h^2$. In consequence, this leads to an increase for somewhat lower densities $\sim 10^6$--$10^7 \, \mathrm{M_\odot} \mathrm{kpc}^{-3} h^2$.
However, the low-density regions ($\lesssim 10^6 \, \mathrm{M_\odot} \, \mathrm{kpc}^{-3} h^2$) do not show any differences between the DM models.

The densest regions are most sensitive to DM self-interactions because the effect of self-interactions depends on the density and the velocity dispersion, which tends to be high in dense regions.
However, we do not find large differences between rSIDM and fSIDM, especially for the smaller cross-section of $\sigma_\mathrm{\Tilde{T}}/m_\chi = 0.1 \, \mathrm{cm}^2 \, \mathrm{g}^{-1}$, the models are very similar.

\subsection{Two-point correlation function}

\begin{figure*}
    \centering
    \includegraphics[width=\columnwidth]{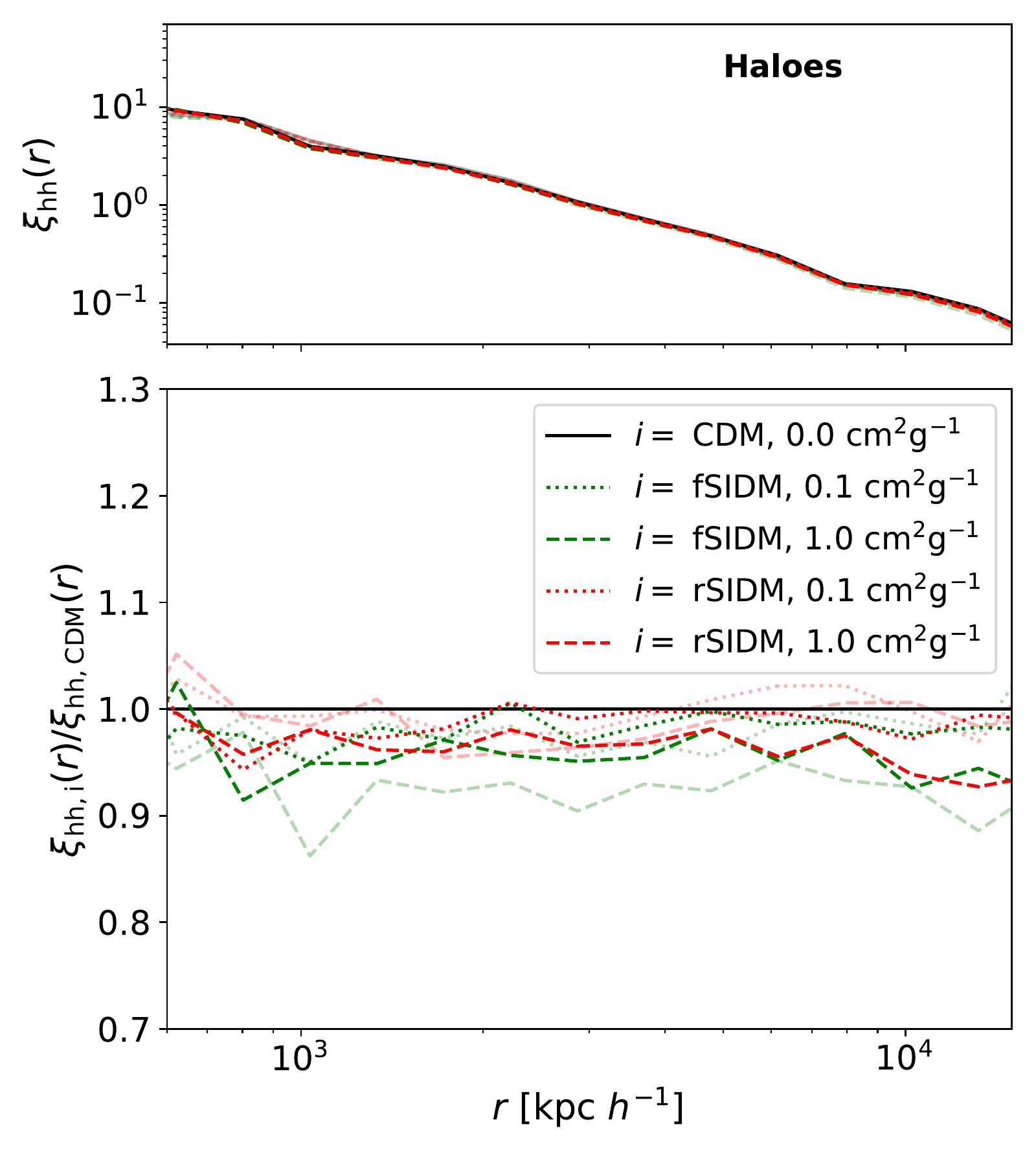}
    \includegraphics[width=\columnwidth]{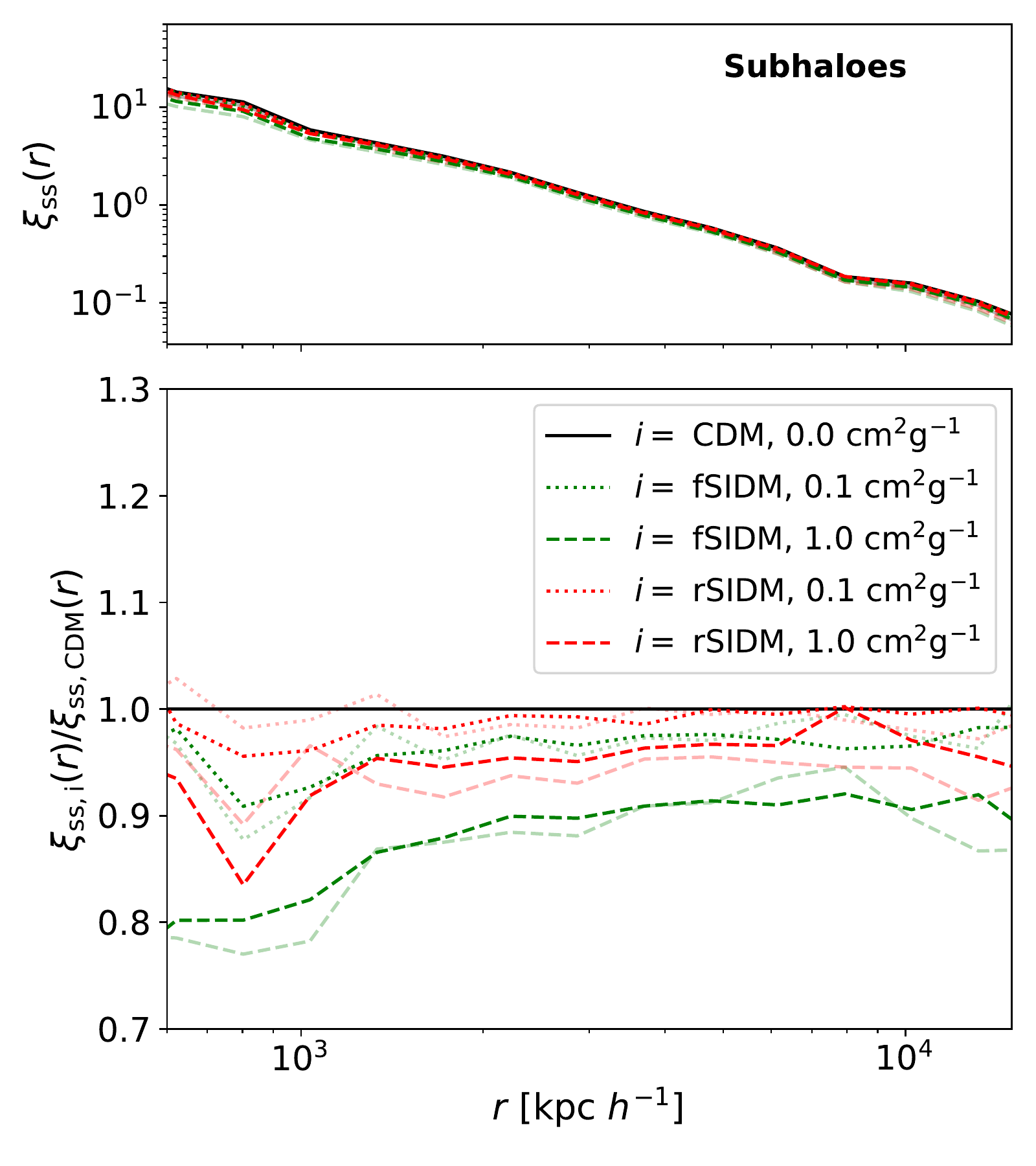}
    \caption{The two-point correlation function is shown for the higher resolution boxes (uhr, darker lines) and the lower resolution boxes (hr, fainter lines).
    For the left-hand panel, we used the halo positions and computed the halo--halo correlation, $\xi_\mathrm{hh}(r)$.
    Likewise, we used the subhalo positions and computed the subhalo--subhalo correlation, $\xi_\mathrm{ss}(r)$, as displayed in the right-hand panel.
    Note that we only considered haloes and subhaloes that have at least a mass of $\sim 9.6 \times 10^{10} \, \mathrm{M_\odot} \, h^{-1}$.
    Results for various DM models, as indicated in the legend, are shown for a redshift of $z=0$.}
    \label{fig:tpcf_halo}
\end{figure*}

In addition to the power spectrum, we can use the two-point correlation function to characterize the distribution of matter. Specifically, we compute the spatial two-point correlation function according to \citep{Davis_1983}
\begin{equation}
    \xi(r) = \frac{N_R}{N_D} \frac{DD(r)}{DR(r)} - 1 \, .
\end{equation}
We use the data points given by the simulation and draw random numbers to generate points from a uniform PDF.
The number of data points is given by $N_D$, and $N_R$ denotes the number of randomly distributed points.
We compute the number of distances $DD(r)$ between data points within the interval $[r, r + \Delta r]$ as well as $DR(r)$, the number of distances between data points and random points.

We compute the halo--halo correlation, $\xi_\mathrm{hh}(r)$, based on the halo positions and the subhalo--subhalo correlation, $\xi_\mathrm{ss}(r)$, based on the subhalo positions as identified by \textsc{subfind} for the full cosmological box simulations.
But we do not use all substructures; instead, we introduce a mass/resolution cut to avoid our results being affected by numerical artefacts. All systems with a mass of $M < 9.6 \times 10^{10} \, \mathrm{M_\odot} \, h^{-1}$ are excluded as they are poorly resolved.
For the uhr runs, this mass corresponds to 2200 particles and for the hr runs to 220 particles.
As a consequence, we have much fewer positions (very roughly $3 \times 10^3$ for hr and uhr as well as haloes and subhaloes) for the computation of the two-point correlation function.
In Fig.~\ref{fig:tpcf_halo}, we show the results for the halo positions (left-hand panel) and the subhaloes (right-hand panel).
For the haloes, we do not find much of a difference among the DM models.
In contrast, the comparison of the subhalo positions reveals a difference on small scales as well as larger scales.
As expected, the stronger the cross-section, the more structures on small scales are suppressed.
For $\sigma_\mathrm{\Tilde{T}}/m_\chi = 1.0 \, \mathrm{cm}^2 \, \mathrm{g}^{-1}$, fSIDM simulations deviate more from CDM than the rSIDM simulations do, with the subhalo correlation function being roughly $20\%$ lower on small scales for fSIDM than CDM.
Note that these results depend on the chosen mass/resolution cut for subhaloes. The lower the cut, the larger the deviations between the CDM and SIDM runs are.

The suppression of the subhaloes compared to haloes may arise from the fact that some of them are satellites and not exclusively primary subhaloes (see Sec.~\ref{sec:numerical_setup}).
The SIDM satellites could dissolve faster because they have lower central densities due to DM self-interactions. Hence, they are not as strongly bound as their CDM counterparts, which makes them more prone to tidal effects. This could be enhanced by scattering between host and satellite particles.
In the next section, we turn to the halo and subhalo mass function.

\subsection{Halo and subhalo mass function} \label{sec:halo_mass_func}

\begin{figure*}
    \centering
    \includegraphics[width=\columnwidth]{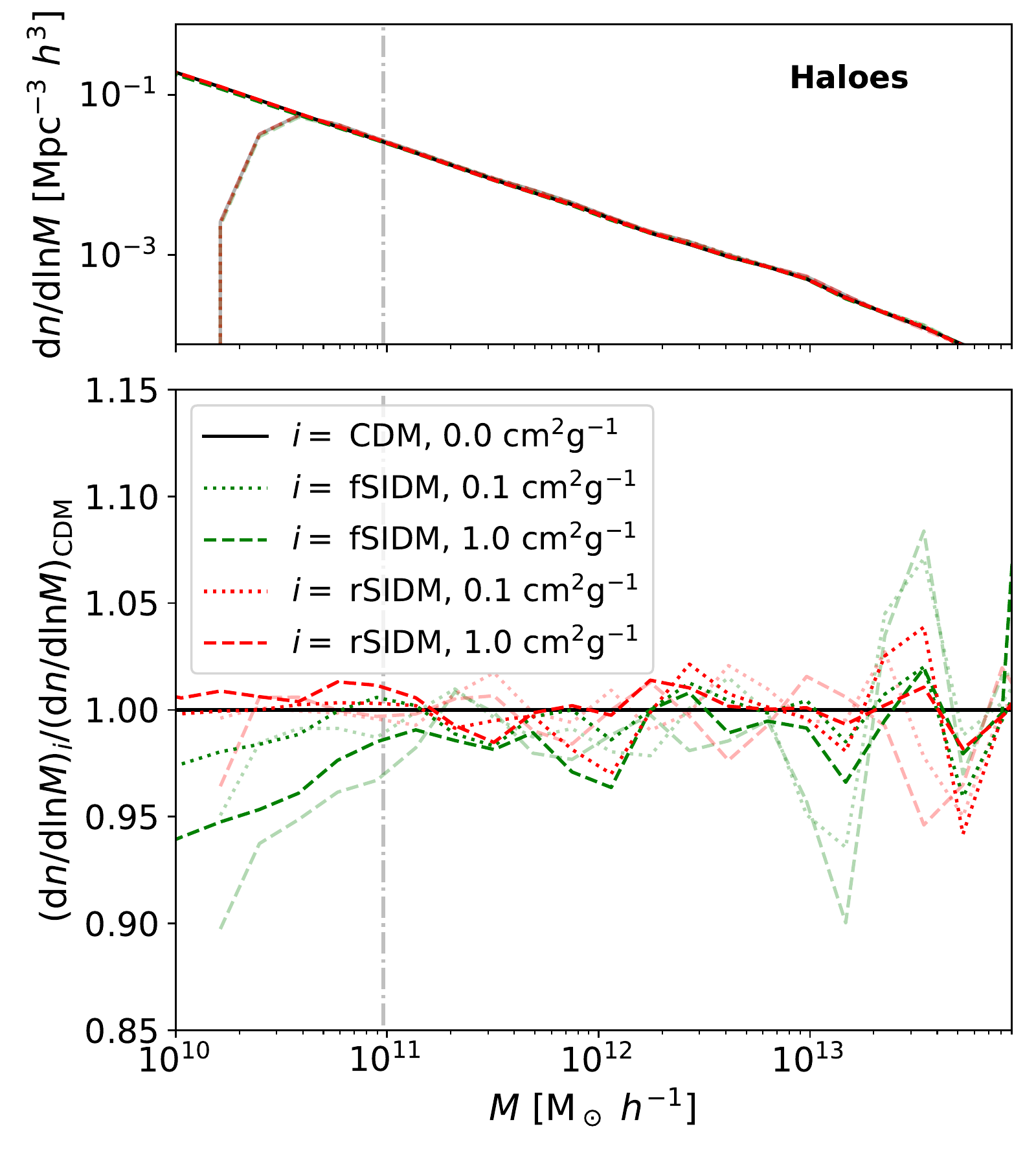}
    \includegraphics[width=\columnwidth]{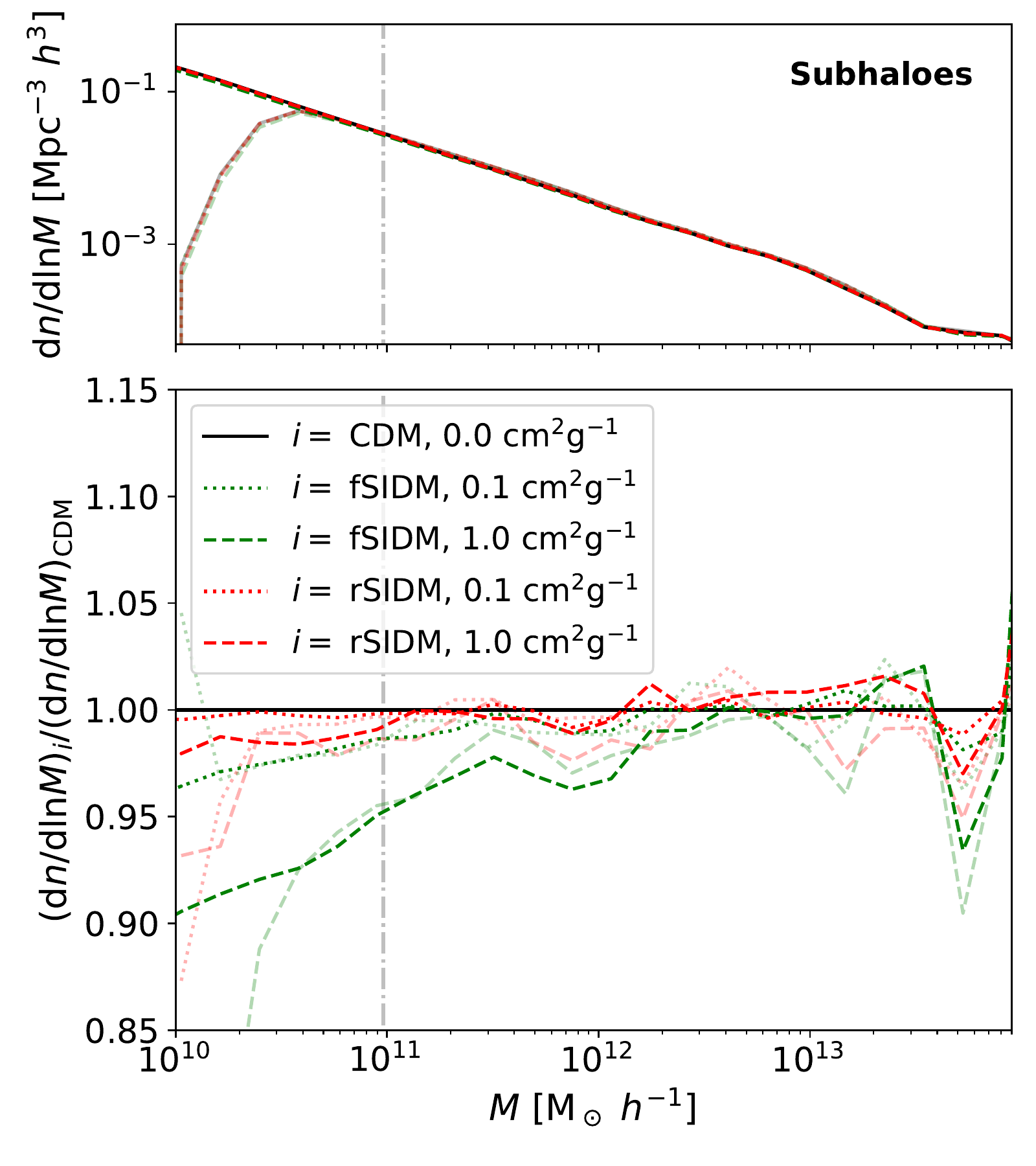}
    \caption{The halo mass (left-hand panel) and the subhalo mass (right-hand panel) function are shown for various simulations.
    We plot the number density of haloes/subhaloes per logarithmic mass bin as a function of the total halo/subhalo mass as identified by \textsc{subfind}.
    The colours indicate the type of the self-interaction and the line style gives the strength of self-interaction as indicated in the legend.
    We display results of the higher resolution boxes (uhr, darker lines) and the lower resolution boxes (hr, fainter lines).
    The dash--dotted grey line indicates the mass limit of $\sim 9.6 \times 10^{10} \, \mathrm{M_\odot} \, h^{-1}$ that we applied previously for the two-point correlation function.
    The plots are for a redshift of $z=0$.}
    \label{fig:mass_func}
\end{figure*}

\begin{figure*}
    \centering
    \includegraphics[width=\columnwidth]{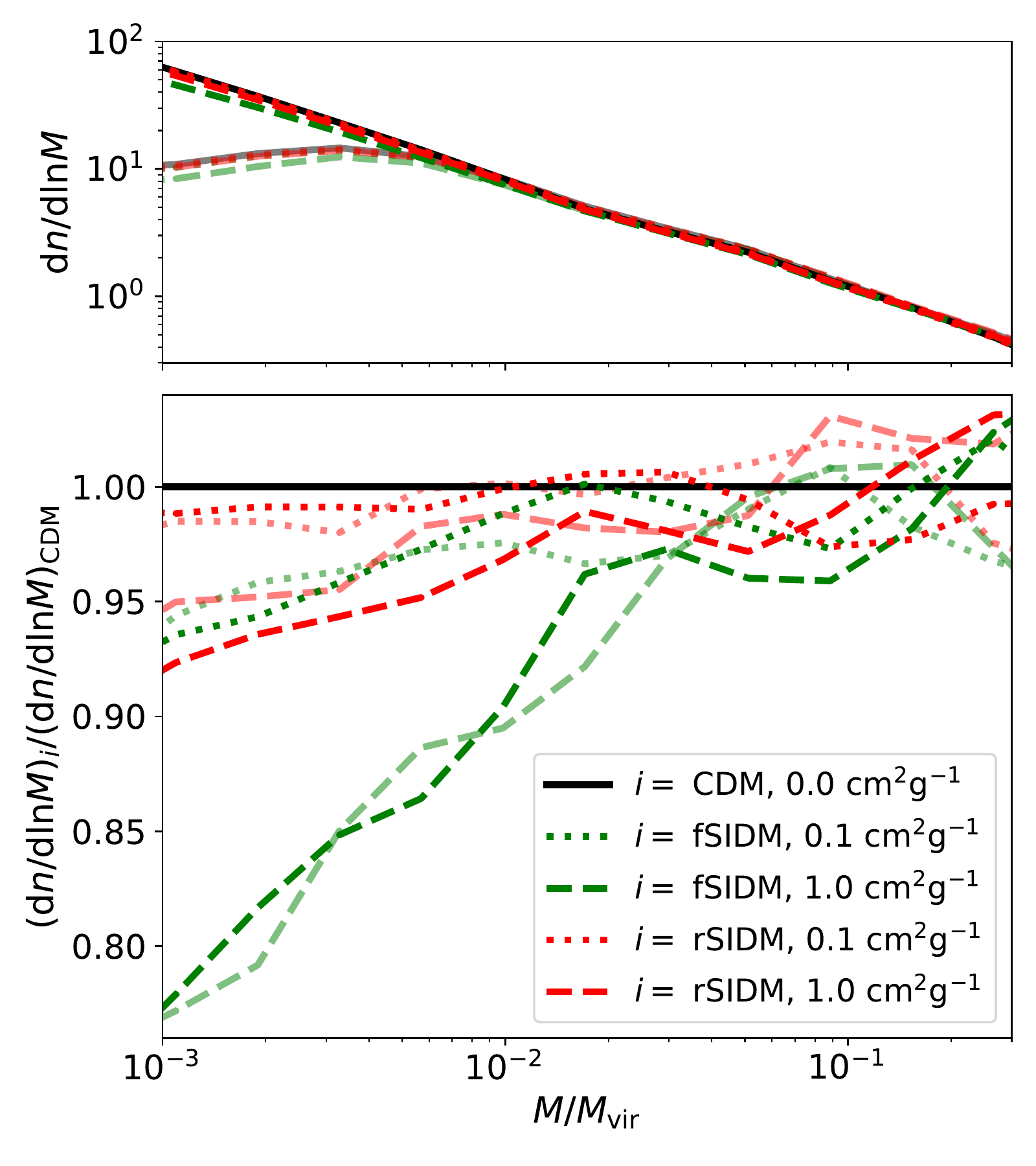}
    \includegraphics[width=\columnwidth]{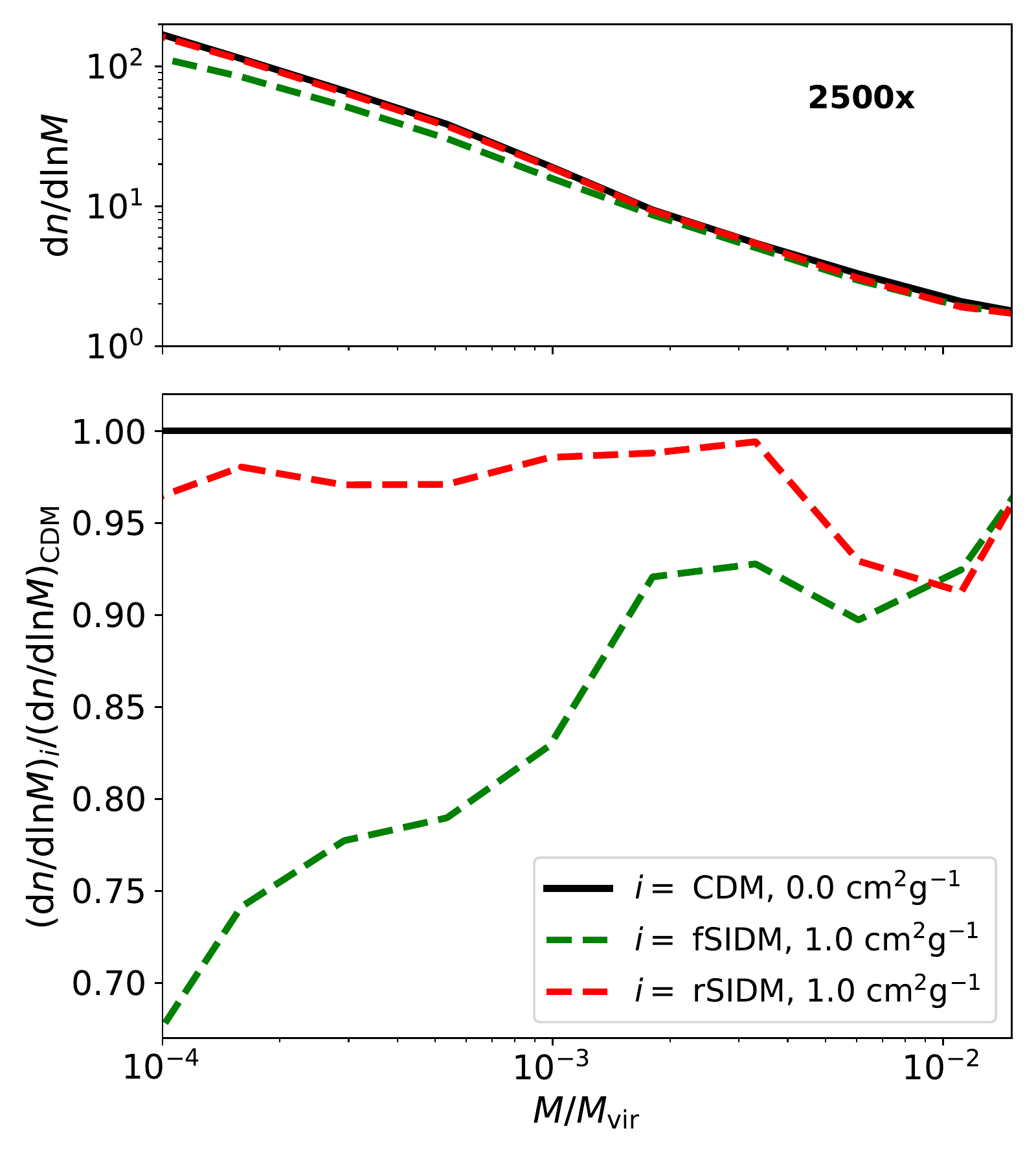}
    \caption{The number of satellites per logarithmic mass as a function of their total mass relative to the virial mass of their host.
    The left-hand panel gives the result of the 100 most massive groups in our full cosmological box with the highest resolution and the lower resolution run (transparent).
    The right-hand panel gives the same but for the three most massive objects in the best resolved zoom-in simulation.
    All subhaloes, except for the primary one, that are within a radius of $5 \, r_\mathrm{vir}$ were considered satellites.
    The results are for a redshift of $z=0$.
    Note that the least resolved satellites of the uhr and x2500 simulations used here contain about 100 particles.
    }
    \label{fig:halo_shmf}
\end{figure*}

\begin{figure}
    \centering
    \includegraphics[width=\columnwidth]{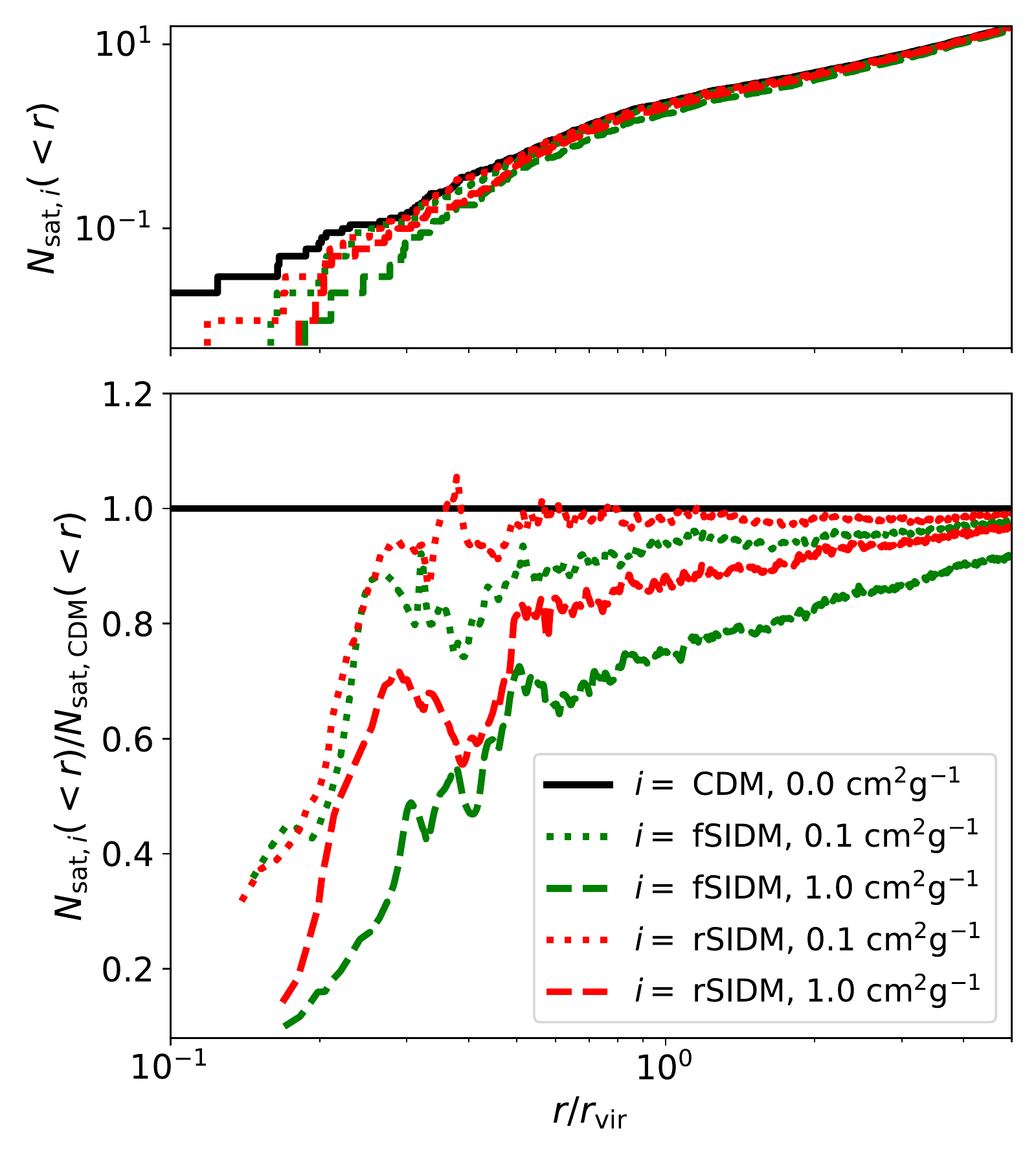}
    \caption{The average cumulative number of satellites per halo as a function of radius is shown for the uhr simulations at $z=0$ (upper panel) as well as the ratio of the DM models to CDM (lower panel). The latter one has been smoothed a little. We use the subhaloes of the 100 most massive haloes and consider them satellites if they are more massive than $M > 9.6 \times 10^{10} \, \mathrm{M_\odot} \, h^{-1}$ and less massive than the primary subhalo.}
    \label{fig:mvd_hist}
\end{figure}

Here, we study the halo and subhalo mass function as well as the abundance of satellites.

In Fig.~\ref{fig:mass_func}, we show the halo mass function (left-hand panel) and the subhalo mass function (right-hand panel).
For the computation, we used the total mass of the haloes and subhaloes calculated by \textsc{subfind}.
At the high-mass end, we do not find a significant difference between the CDM and SIDM models.
But at lower masses, self-interactions suppress the number of haloes and even more the number of subhaloes. However, the smallest objects we display here are not well resolved.
Note that the mass cut previously used for Fig.~\ref{fig:tpcf_halo} selects systems with a mass $\gtrsim 9.6 \times 10^{10} \, \mathrm{M_\odot} \, h^{-1}$.
At least the largest fSIDM cross-section studied here gives us a significant reduction of the number of subhaloes above this cut.
But for the haloes, none of the cross-sections results in a significant reduction for masses $\gtrsim 9.6 \times 10^{10} \, \mathrm{M_\odot} \, h^{-1}$.
The difference between the halo and subhalo mass function can only arise from satellites dissolving faster, as all non-satellites are haloes.
Hence, satellites, in particular low-mass satellites, appear to be an interesting test bed for SIDM models.
Several SIDM studies have focused on them \citep[e.g.][]{Kahlhoefer_2019, Kaplinghat_2019a, Banerjee_2020, Nadler_2020, Nadler_2021, Nishikawa_2020, Sameie_2020, Yang_2020, Correa_2021, Zeng_2021}. 

In Fig.~\ref{fig:halo_shmf}, we study the abundance of satellites for the DM models.
For each group, we compute the number of satellites per logarithmic mass as a function of the satellite mass divided by the virial mass of the group.
We select all subhaloes, except the primary one, within $5 \, r_\mathrm{vir}$. The results for a selection criterion of $1 \, r_\mathrm{vir}$ are shown in Appendix~\ref{sec:shmf2}.
For the full-box simulations, we do this for the 100 most massive groups and take the mean (left-hand panel).
In contrast, for the zoom-in simulations, we only take the mean of the three most massive systems (right-hand panel).
The comparison between SIDM and CDM in the lower panel shows that self-interactions can suppress the abundance of satellites.
The stronger the self-interactions, the fewer satellites are present, and low-mass satellites are more affected than more massive ones.
However, we have to note that the lowest mass satellites used in Fig.~\ref{fig:halo_shmf} are not well resolved.
Although the halo mass function is not converged for the hr run at low masses (left-hand panel), the differences to CDM, in particular for fSIDM with $\sigma_\mathrm{\Tilde{T}}/m_\chi = 1.0 \, \mathrm{cm}^2 \, \mathrm{g}^{-1}$, seem to be converged.
This makes it plausible that the effect for the lower masses is real.
In a similar manner, \cite{Stafford_2020} found the difference between cosmologies to be converged for density profiles below the convergence radius proposed by \cite{Ludlow_2019}.
We find differences for better resolved haloes at larger masses too, but not for the most massive satellites.
For the larger cross-section of $\sigma_\mathrm{\Tilde{T}}/m_\chi = 1.0 \, \mathrm{cm}^2 \, \mathrm{g}^{-1}$, frequent self-interactions seem to be much more efficient in reducing the number of satellites than rare scatterings.
This is in line with the suppression of the spatial two-point correlation function of subhaloes at small scales (right-hand panel of Fig.~\ref{fig:tpcf_halo}).
It is also worth mentioning, which we found earlier, in a study of head-on collisions of unequal-mass mergers \citep{Fischer_2021b}, that fSIDM subhaloes dissolve faster than rSIDM subhaloes when they are matched in terms of $\sigma_\mathrm{\Tilde{T}}/m_\chi$.
We have to note that there cannot exist a perfect matching between rare and frequent self-interactions as they are qualitatively different.
The $\sigma_\mathrm{\Tilde{T}}$-matching can only provide an approximation that can be helpful in some situations, e.g.\ density profile (see Sec.~\ref{sec:density_profiles}) or shape profiles (see Sec.~\ref{sec:shapes}).
In Sec.~\ref{sec:fSIDMvsrSIDM}, we further investigate the qualitative difference concerning the abundance of satellites.

In addition, we study the radial dependence of the suppression of the satellite abundance in Fig.~\ref{fig:mvd_hist}.
Here, we find that the effect of SIDM on the number of satellites becomes stronger at smaller distances to the host.
As in Figs~\ref{fig:mass_func} and \ref{fig:halo_shmf}, we find that fSIDM reduces the abundance of satellites stronger than rSIDM given the same value for $\sigma_{\tilde{T}} / m_\chi$.
However, we have to note that for small radii the number of satellites we study here is low and thus the error on the ratios shown in Fig.~\ref{fig:mvd_hist} is sizeable. Thus, we can only make a qualitative statement for the smallest radii, but not quantify the difference between the DM models.

That satellites can dissolve faster in the presence of SIDM has been found earlier.
The evolution of a satellite and its lifetime depends on several aspects and physical mechanisms that are at play.
In CDM, satellites are torn apart by tidal forces, which act against their gravitational self-binding.
The more massive and the more concentrated a satellite is, the more resistant it is against the tidal forces and can survive longer.
In the context of SIDM, a DM core can form, which flattens the gravitational potential and makes the satellite more prone to tidal disruption \citep{Yang_2020}.
However, if the satellite is in a further evolution phase and undergoes core collapse \citep[e.g.][]{Balberg_2002, Koda_2011, Essig_2019}, it is more protected against tidal disruption as it is even more concentrated than its CDM counterparts.
\cite{Nishikawa_2020} found that tidal stripping enhances the core collapse in SIDM haloes, i.e.\ reduces the collapse time.
A recent study \citep{Zeng_2021} found that it is nearly impossible for satellites to undergo core collapse if self-interactions are velocity-independent.
The evolution of a satellite is not only determined by tidal stripping and DM self-interactions between satellite particles, it is also affected by tidal heating and DM self-interactions between satellite and host particles.
Whether tidal effects (stripping and heating) enhance or prevent core collapse depends on the mass concentration of the satellite.
The satellite--host interactions transfer energy into the satellite and thus contribute to the core formation.
For a cross-section, which is decreasing with velocity \citep[e.g.][]{Loeb_2011, Tullin_2012}, this effect would become much less as the typical relative velocity between satellite and host particles is larger than that between satellite particles.
In consequence, the survival time of satellites is expected to depend crucially on the velocity dependence of the self-interactions.
\cite{Banerjee_2020} studied various SIDM models and found that the suppression of the satellite abundance is weaker if the cross-section is velocity-dependent.
However, they also studied an anisotropic cross-section and did not find any significant deviation to an isotropic cross-section, and this also includes satellite counts.
It is worth pointing out that our fSIDM cross-section is more anisotropic.
In practice, it might be infeasible to study it with an rSIDM scheme, as used by \cite{Banerjee_2020}, because the high scattering rate would require very small time-steps.
In addition, we should mention the work by \cite{Vogelsberger_2019}.
They studied an MW-like halo with multistate inelastic DM self-interactions and found that this type of interaction suppresses the abundance of small structures even for small cross-sections considerably.

Overall, satellites are not only interesting objects that constrain the strength of SIDM, but may also provide bounds on the angular and velocity dependence of the differential cross-section.
In Sec.~\ref{sec:fSIDMvsrSIDM}, we discuss further potential strategies to constrain this angular dependence.

\subsection{Density profiles} \label{sec:density_profiles}

\begin{figure}
    \centering
    \includegraphics[width=\columnwidth]{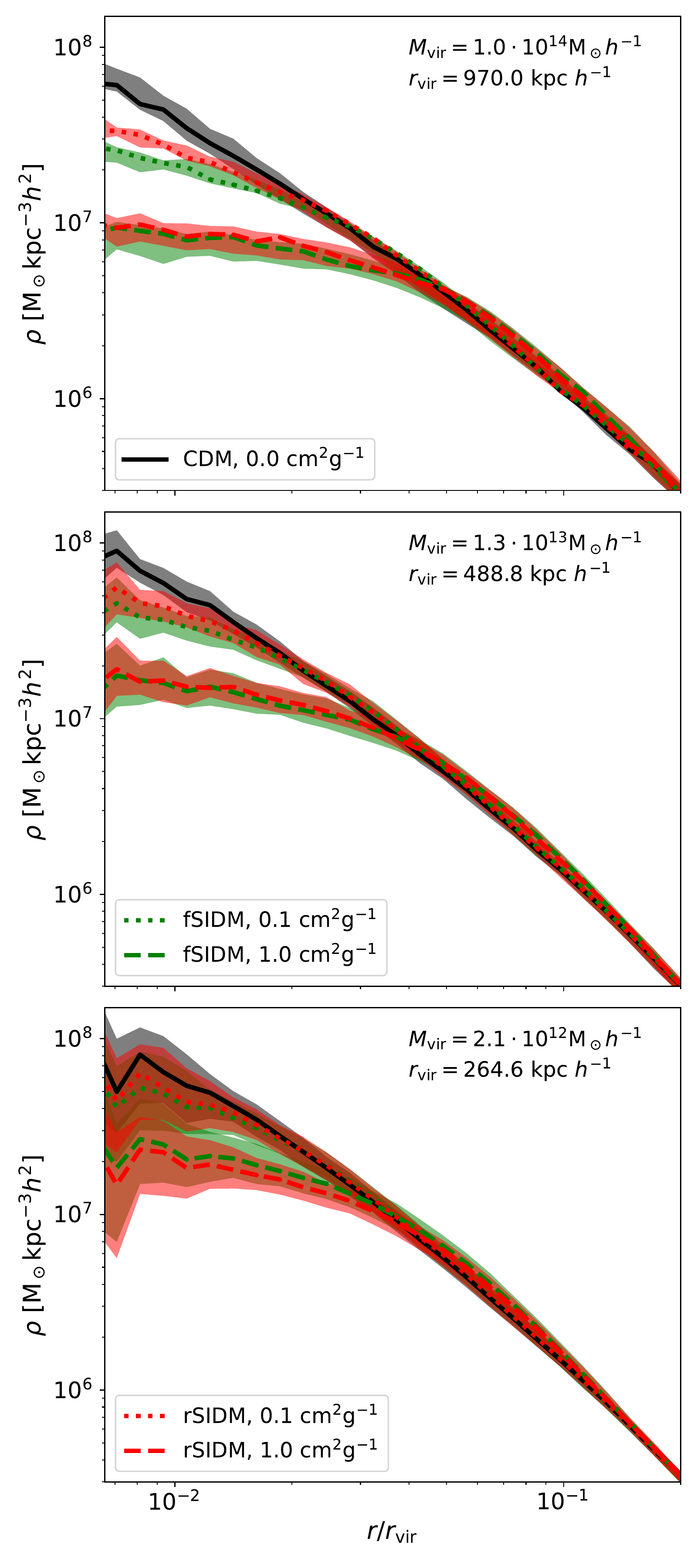}
    \caption{Median density profiles are shown for various halo mass bins and cross-sections.
    The density is plotted as a function of the radius in units of the virial radius.
    The shaded regions indicate the scatter among the haloes, and the range between the 25th and 75th percentiles is displayed.
    The virial mass and the virial radius given in the panels indicate the median of the corresponding mass bin.
    All plots show the profiles for a redshift of $z=0$ and are produced from the full cosmological box with the highest resolution.
    Note that we have used all particles, not only those that belong to the halo as identified by \textsc{subfind}.
    }
    \label{fig:halo_density_profiles}
\end{figure}

\begin{figure}
    \centering
    \includegraphics[width=\columnwidth]{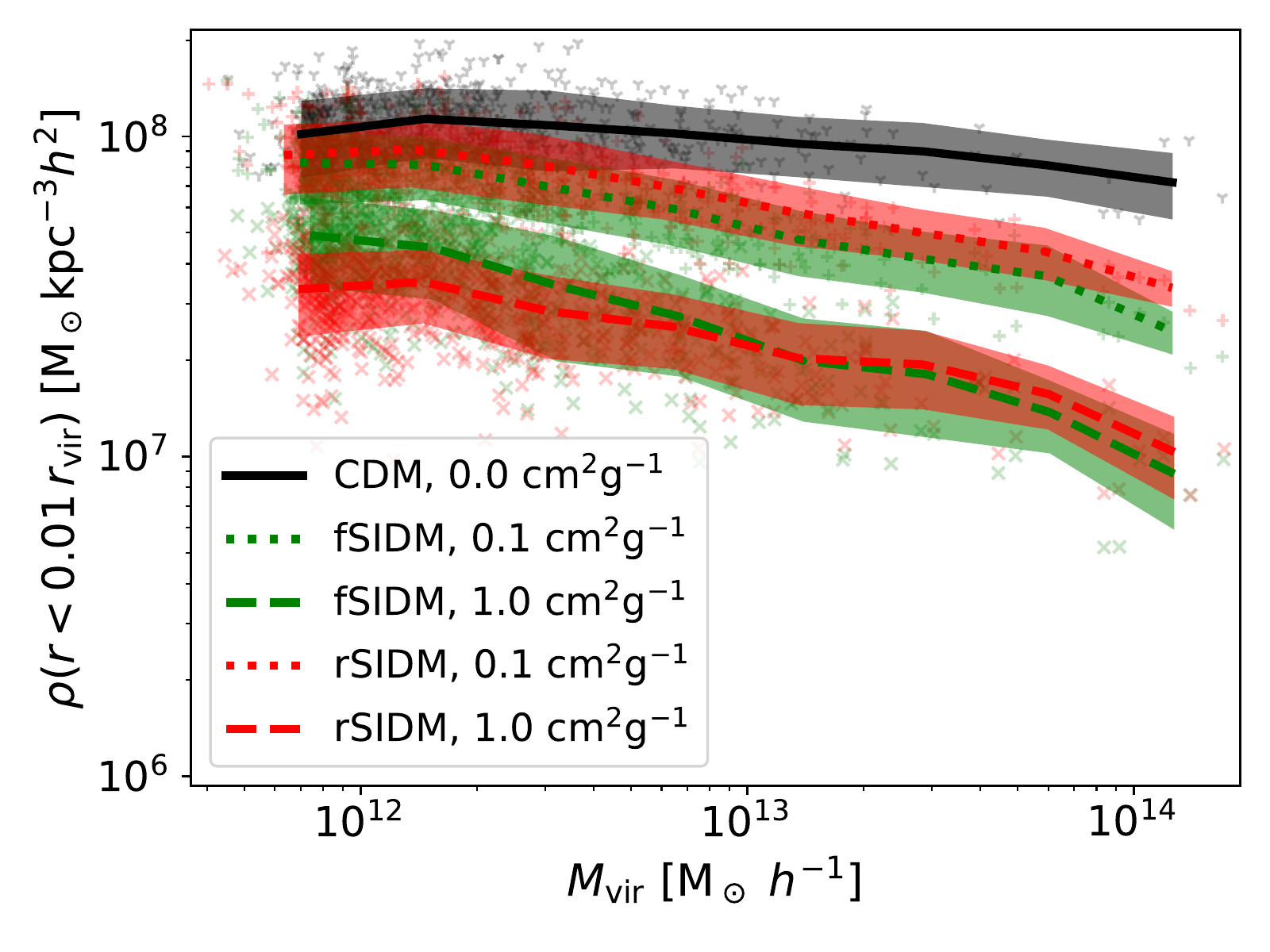}
    \caption{We display the central density of haloes as a function of their virial mass.
    The results of various simulations at a redshift of $z=0$ are shown.
    The central density is measured as the mean density within a sphere of $0.01 \, r_\mathrm{vir}$.
    Systems evolved with the smaller cross-section are marked by ``$+$'' and for the larger cross-section we use ``$\times$''; the CDM case is indicated by ``$\downY$''.
    In addition to the individual systems, we computed the mean of the distribution as a function of virial mass, indicated by the lines.
    The shaded region gives the standard deviation.}
    \label{fig:central_density}
\end{figure}

As we have seen in Fig.~\ref{fig:pdd}, in scenarios with SIDM the high-density regions are suppressed.
In particular, density cores have been studied in the literature and used to constrain self-interactions \citep[e.g.][]{Correa_2021, Sagunski_2021, Ray_2022}. Recent strong lensing observations provide further evidence for DM cores in galaxy clusters \citep{Limousin_2022}.

In cosmological simulations of rSIDM, density and circular velocity profiles of haloes have been studied previously by \cite{Rocha_2013, Robertson_2019, Banerjee_2020}, and \cite{Stafford_2020}.

In Fig.~\ref{fig:halo_density_profiles}, we show median density profiles of the haloes for three different mass bins.
In the outer regions, the density profiles are very similar among the various DM models.
But in the central region, we observe a density core for SIDM while CDM predicts cuspy haloes.
The central density is lower for a larger cross-section and frequent self-interactions lead mostly to slightly larger cores than rare self-interactions, if a $\sigma_\mathrm{\Tilde{T}}$-matching (same momentum-transfer cross-section for rSIDM and fSIDM) is employed.
This is in agreement with previous findings of isolated haloes \citep{Fischer_2021a}.

The central density of the individual haloes is shown in Fig.~\ref{fig:central_density}. Specifically, we computed the mean density within $0.01 \, r_\mathrm{vir}$.
As we have already seen above, a larger self-interaction cross-section leads to haloes with lower central densities.
For CDM, we observe a slight decline of the central density with virial mass.
But for the SIDM models, the decline is steeper.
This implies an increasing difference between CDM and SIDM models with virial mass.
Note that with increasing virial mass the velocity dispersion in the inner regions of the DM haloes increases and thus the self-interactions are more efficient.

\subsection{Circular velocity}

\begin{figure}
    \centering
    \includegraphics[width=\columnwidth]{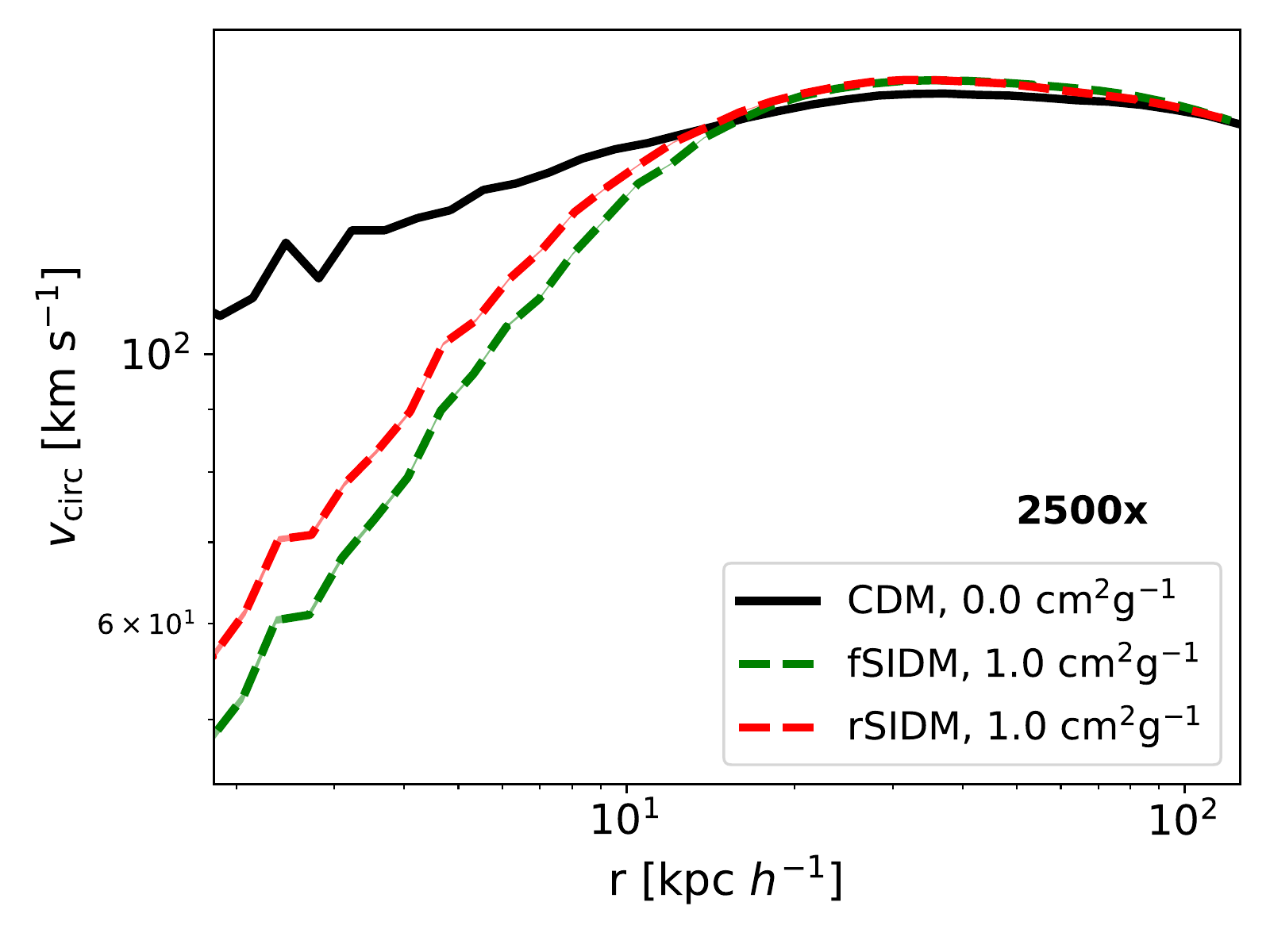}
    \caption{The circular velocity as a function of radius is shown for the most massive subhalo of our highest resolved zoom-in simulation at a redshift of $z=0$.}
    \label{fig:subhalo_vcirc_profile}
\end{figure}

\begin{figure}
    \centering
    \includegraphics[width=\columnwidth]{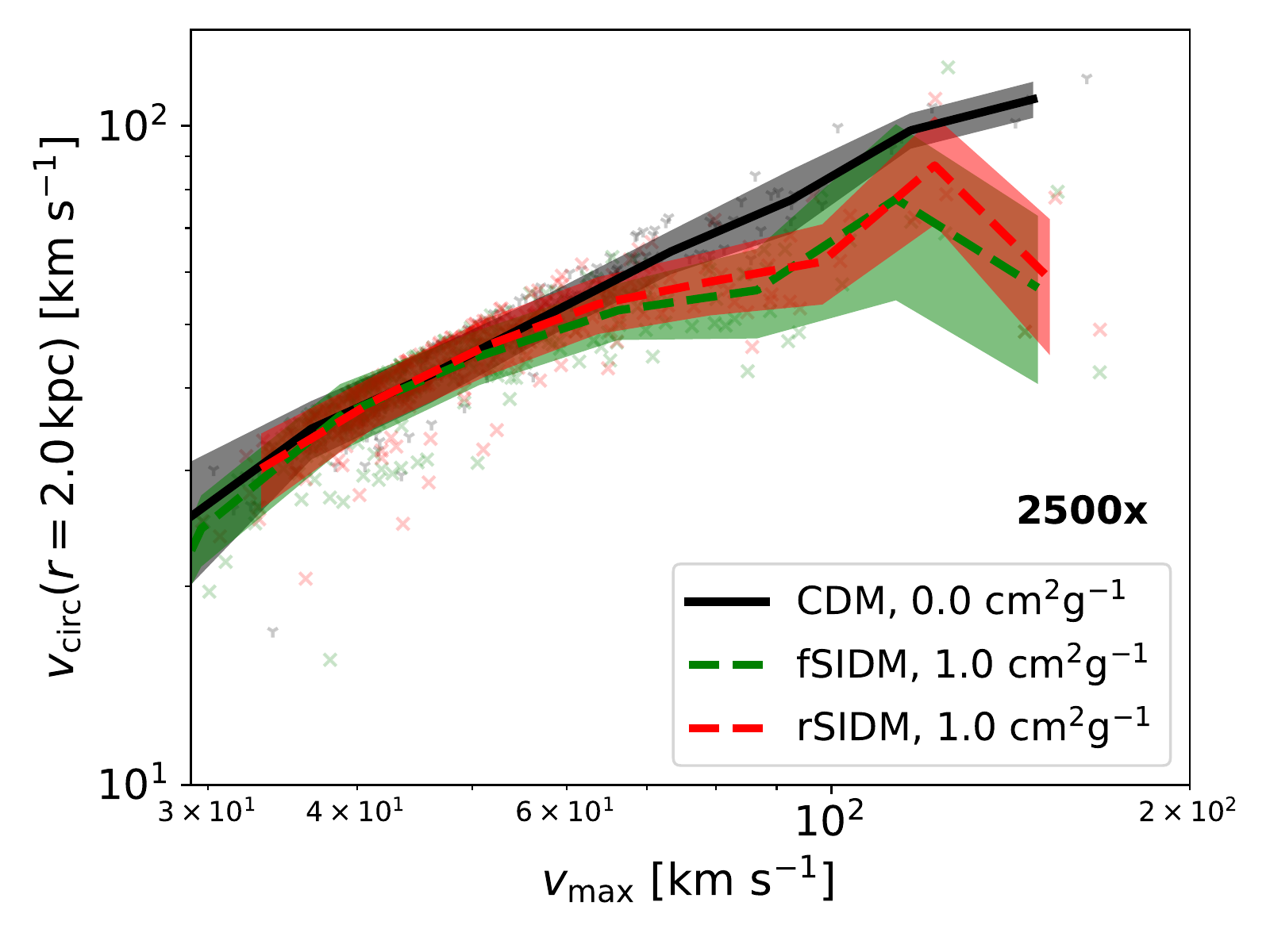}
    \caption{The circular velocity at 2 kpc vs.\ the maximum circular velocity measured from the circular velocity profile as shown in Fig.~\ref{fig:subhalo_vcirc_profile} is displayed.
    Each dot corresponds to one of the 300 most massive subhaloes of the highest resolved zoom-in simulation. The lines correspond to the mean and the shaded regions indicate the standard deviation.}
    \label{fig:vcirc_vmax}
\end{figure}

In addition to computing the density profile, we can also study the circular velocity,
\begin{equation} \label{eq:vcirc}
    v_\mathrm{circ} (r) = \sqrt{\frac{\mathrm{G} M(<r)}{r}} \,,
\end{equation}
as a function of the radius.
In Fig.~\ref{fig:subhalo_vcirc_profile} we show the circular velocity profile for the most massive subhalo of the best resolved zoom-in simulation.
It is visible that in the inner regions the velocity that is needed for a circular orbit is less for SIDM compared to CDM.
This is a direct consequence of the density core, i.e.\ the enclosed mass, $M(<r)$, in Eq.~\eqref{eq:vcirc} is less.
At larger radii, the circular velocity is the same across the DM models.

In order to trace the core formation in terms of $v_\mathrm{circ}$, we can measure it at a small radius and compare it to its maximum value.
We do so in Fig.~\ref{fig:vcirc_vmax} and plot $v_\mathrm{circ}(2 \, \mathrm{kpc})$ vs.\ $v_\mathrm{circ, max}$ for the 300 most massive subhaloes of the best resolved zoom-in simulation.
For the most massive subhaloes, i.e.\ the ones with a larger value for $v_\mathrm{max}$, we find a difference between the DM models.
The SIDM models tend to have lower circular velocities at 2 kpc than CDM.
For fSIDM, $v_\mathrm{circ}$ is slightly lower than for rSIDM, which is in line with the lower densities found for fSIDM.
At lower masses, we do not find a significant difference; here a measure at 2 kpc may not be sensitive to the density core.
Instead, a measure at smaller radii would be preferable to trace the density core.

Self-interactions have turned out to be interesting to explain the diversity of observed rotation curves \citep[e.g.][]{Oman_2015}.
In particular, the response of SIDM to the gravitational potential of the baryons can increase the diversity of rotation curves \citep[e.g.][]{Creasey_2017, Kamada_2017, Kaplinghat_2019a}.
However, \cite{Zentner_2022} claim that it is not clear whether observed galactic rotation curves are better explained by SIDM or CDM including baryons.
For late-type dwarfs \cite{Roper_2022} points out that the measured rotation curves could differ significantly from their circular velocity curves.
Other studies had also previously indicated that the kinematic modelling of these objects might be problematic \citep[e.g.][]{Pineda_2016, Read_2016b, Genina_2018, Oman_2019}.
In any case, this question can only be addressed once baryonic physics is taken into account in the $N$-body code, implying that our results cannot be directly compared to these findings.
We leave a more in-depth discussion of these issues for future work.

\subsection{Shapes} \label{sec:shapes}

\begin{figure}
    \centering
    \includegraphics[width=\columnwidth]{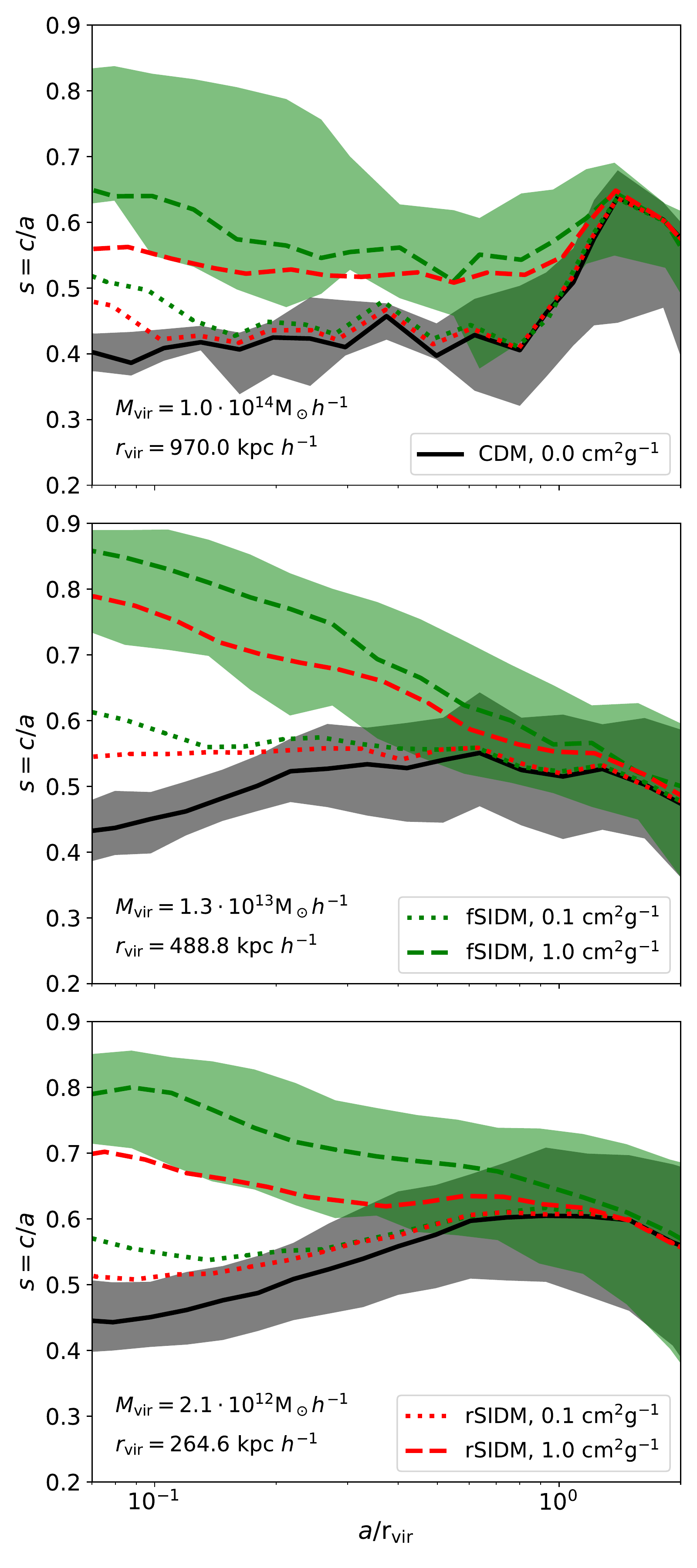}
    \caption{The median halo shapes for different mass bins are shown as a function of the median semimajor axis in units of the virial radius.
    We compute the median shape as well as the median semimajor axis from ellipsoids having the same volume.
    The shaded regions indicated the scatter among the haloes, and the range between the 25th and 75th percentiles is displayed.
    For the sake of clarity, we show this only for the CDM simulation and the fSIDM run with the larger cross-section.
    The virial mass given in the panels indicates the median virial mass of the haloes of the corresponding mass bin.
    The same applies to the shown virial radius.
    Note that we are using the same mass bins as in Fig.~\ref{fig:halo_density_profiles}.
    The shapes are computed from the highest resolved full cosmological box and for a redshift of $z=0$.
    }
    \label{fig:halo_shape_s}
\end{figure}

\begin{figure}
    \centering
    \includegraphics[width=\columnwidth]{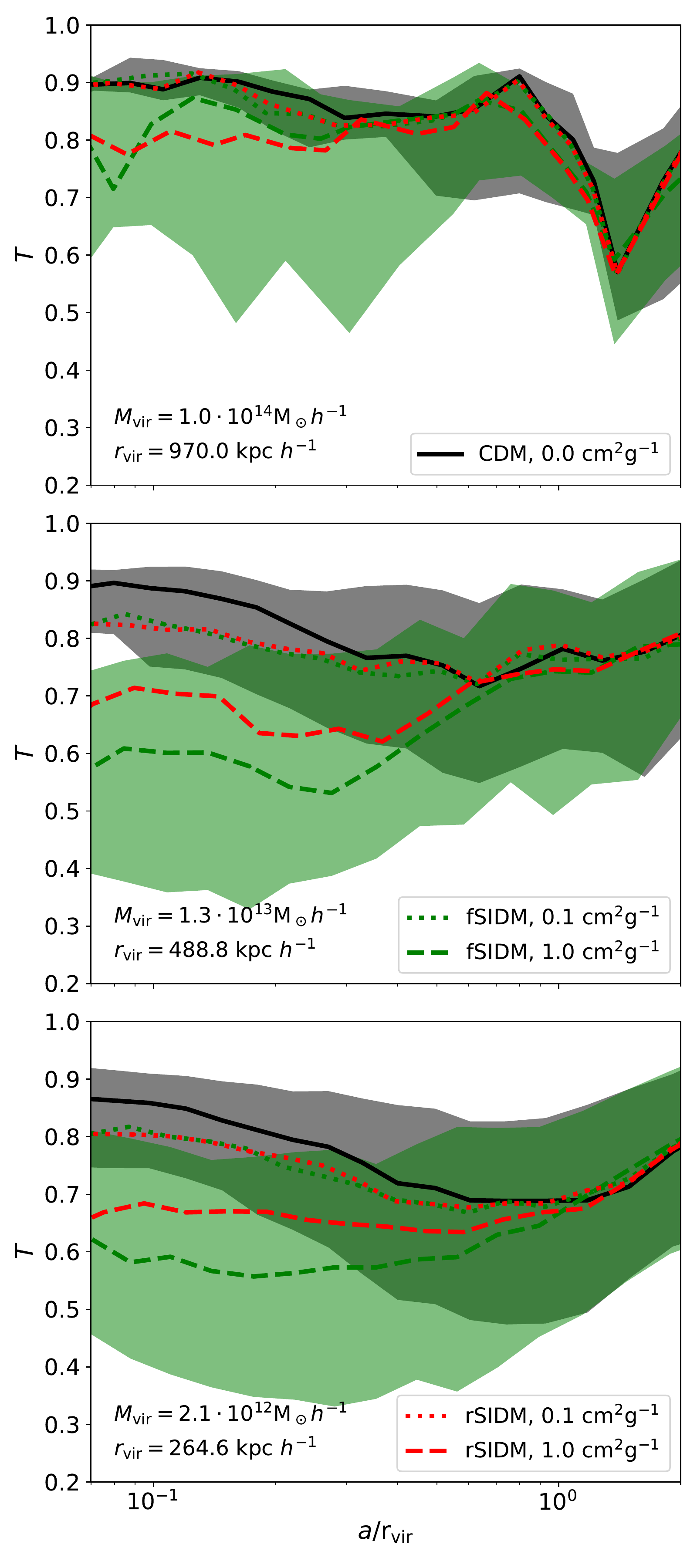}
    \caption{The same as in Fig.~\ref{fig:halo_shape_s} but we show the triaxiality $T$ instead of $s=c/a$. The triaxiality is computed according to Eq.~\eqref{eq:triaxiality}.}
    \label{fig:halo_shape_t}
\end{figure}

\begin{figure}
    \centering
    \includegraphics[width=\columnwidth]{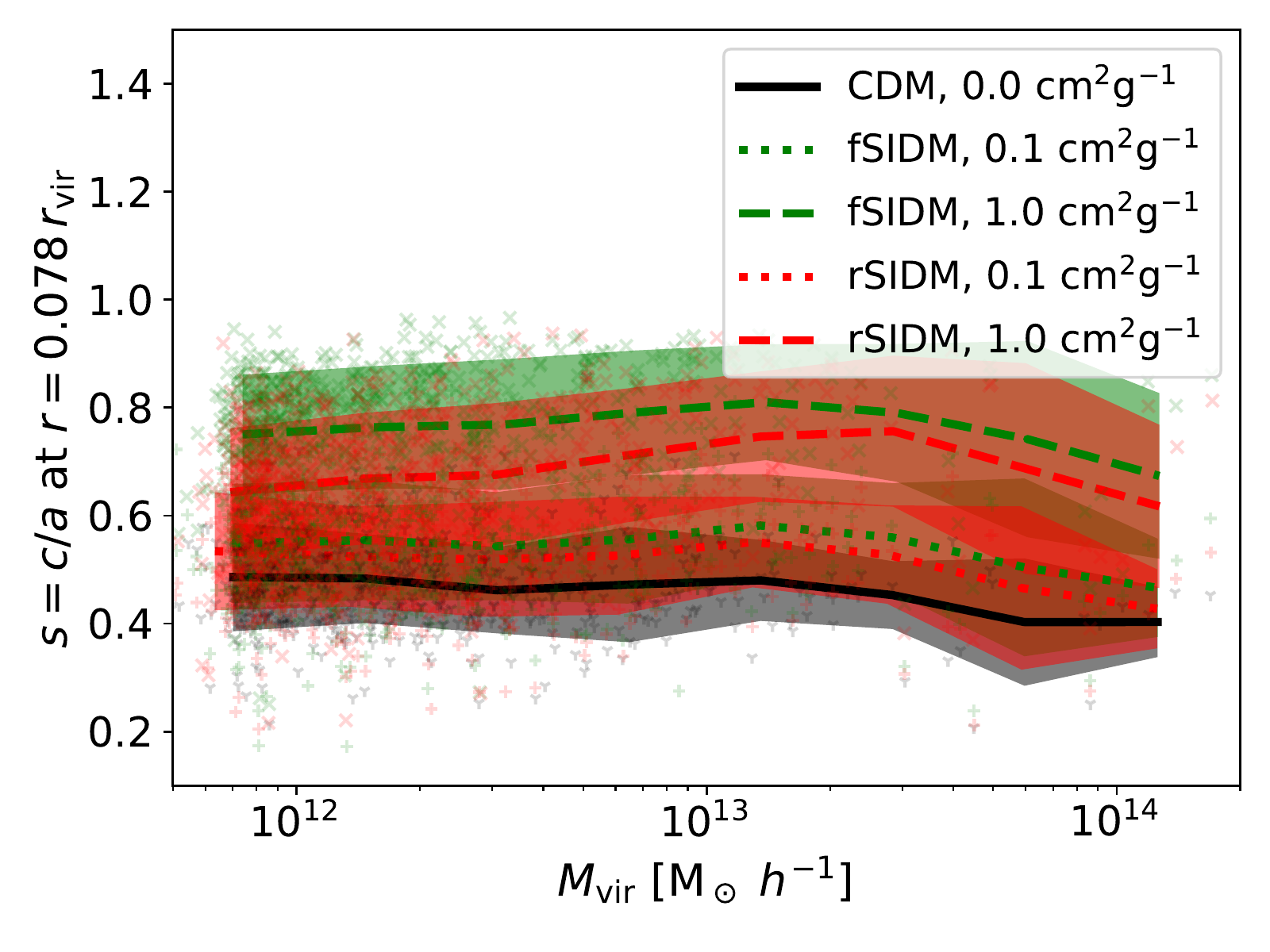}
    \caption{The shape of individual haloes as a function of their virial mass at a redshift of $z=0$ is shown.
    The shapes are measured as $s = c/a$ within an ellipsoid that has the volume of a sphere of $0.078 \, r_\mathrm{vir}$.
    Systems evolved with the smaller cross-section are marked by ``+'' and for the larger one we use ``$\times$''; the CDM case is indicated by ``$\downY$''.
    In addition to individual systems, we computed the mean of the distribution as a function of virial mass, indicated by the lines.
    The shaded region gives the standard deviation.}
    \label{fig:shape_s}
\end{figure}

The shape of DM haloes can provide bounds on the strength of self-interactions. 
We use the tensor, $\mathbfit{S}$, to compute the shape. For $N$ point masses, it is
\begin{equation} \label{eq:shape_tensor}
    \mathbfit{S} \equiv \frac{\sum^N_{n=1} m_n\,\mathbfit{r}_n\otimes\mathbfit{r}_n}{\sum^N_{n=1} m_n} \,,
\end{equation}
with the particle mass $m_n$ at position $\mathbfit{r}_n$.
In order to compute the shapes, we use the eigenvalues ($\lambda_1 \geq \lambda_2 \geq \lambda_3$). They are related to the semi-axes ($a \geq b \geq c$).
In particular, we measure the axial ratios $s = c/a = \sqrt{\lambda_3/\lambda_1}$ and $q = b/a = \sqrt{\lambda_2/\lambda_1}$.
Based on these two ratios, we compute the triaxiality \citep{Franx_1991},
\begin{equation} \label{eq:triaxiality}
    T \equiv \frac{a^2 - b^2}{a^2 - c^2} = \frac{1 - q^2}{1 - s^2} \, .
\end{equation}
Our method to measure halo shapes is iterative, where during each iteration the computation of the tensor, $\mathbfit{S}$, uses all particles within an ellipsoid. The orientation and axial ratios of this ellipsoid are based on $\mathbfit{S}$ from the previous iteration, and the volume of the ellipsoid is kept constant throughout the iterations. We iterate until the two axial ratios ($s$ and $q$) are converged.
Note that in the literature various methods to measure the shape of haloes have been used \citep[e.g.][]{Zemp_2011, Peter_2013, Sameie_2018, Robertson_2019, Banerjee_2020, Chua_2020, Harvey_2021, Vargya_2021, Shen_2022}.

In Fig.~\ref{fig:halo_shape_s}, we show the median shapes for three halo mass bins.
Note that for the computation we have used all particles, not only those that belong to the halo as identified by \textsc{subfind}.
The first mass bin contains only a few objects leading to more noise in the upper panel.
At small distances from the centre, the self-interactions lead to rounder haloes, mainly depending on the strength of the self-interactions.
This has been discovered earlier \citep[e.g.][]{Peter_2013}.
The difference between fSIDM and rSIDM is small, with fSIDM leading to haloes that are a little more spherical than the rSIDM haloes, given a $\sigma_\mathrm{\Tilde{T}}$-matching.
At larger radii, when more particles are included, the shapes become very similar between the different DM models.
These additional particles are located at lower densities and are hence hardly affected by self-interactions.
However, the SIDM shapes differ from CDM at radii beyond the density core more than the density profiles do.
We have evaluated this in great detail in Appendix~\ref{sec:dslr}.
Furthermore, we find that the radius where they start to differ from each other depends strongly on the strength of the self-interactions.
This has been previously pointed out by \cite{Vargya_2021} and suggested as a measure of the total cross-section.

In Fig.~\ref{fig:halo_shape_t}, we also show a complementary shape variable, the triaxiality $T$, as given in Eq.~\eqref{eq:triaxiality} for the same mass bins as in Fig.~\ref{fig:halo_shape_s}.
We observe that $T$ is decreasing with increasing cross-section, implying that the haloes become less prolate.
For large radii, where the matter density becomes smaller, differences between the DM models vanish as expected.

We finally show the shape ($s=c/a$) measured in ellipsoids with a volume equal to a sphere with $0.078 \, r_\mathrm{vir}$ of individual systems as a function of the virial mass in Fig.~\ref{fig:shape_s}.
As already expected from the shape profiles, we observe that the haloes become more spherically symmetric with increasing cross-section.
Again, fSIDM leads to rounder haloes than rSIDM.
There is no significant qualitative difference between the trend of rare and frequent scattering with virial mass.
At the high-mass end, we have only very few objects such that the apparent decrease of $s$ may not be significant.

\subsection{fSIDM versus rSIDM} \label{sec:fSIDMvsrSIDM}

\begin{figure}
    \centering
    \includegraphics[width=\columnwidth]{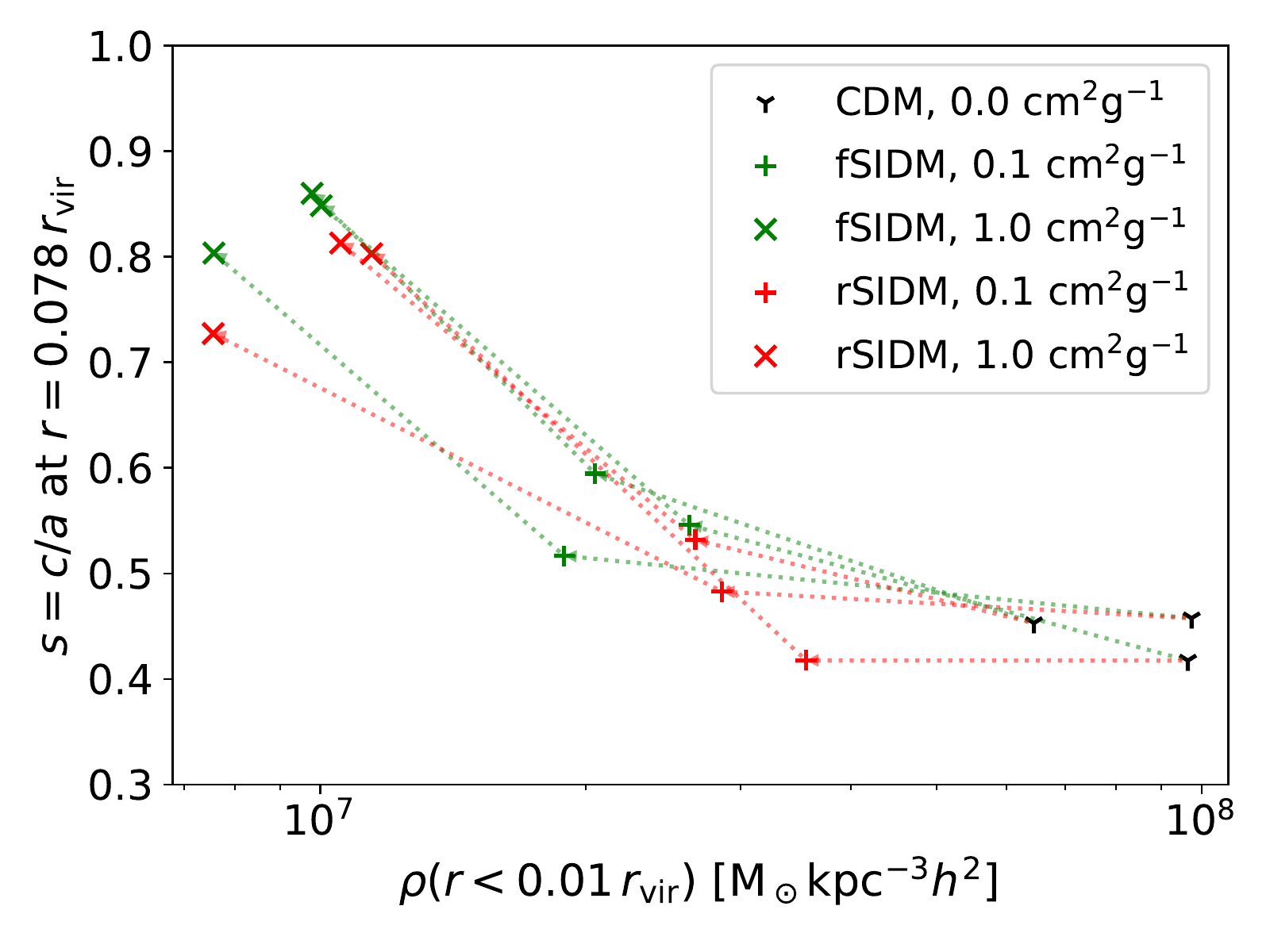}
    \caption{The shape within an ellipsoid with the volume of a sphere with a radius of $r = 0.078 \, r_\mathrm{vir}$ is shown as a function of the central density within a sphere of $r = 0.01 \, r_\mathrm{vir}$ for individual systems.
    The colours indicate the type of self-interaction and the symbols give their strength.
    We show the three most massive haloes ($\sim 1.2$--$1.7 \times 10^{14} \, \mathrm{M_\odot} \, h^{-1}$) of the highest resolved full-box simulation at a redshift of $z=0$.
    How they evolve when increasing the cross-section is indicated by the arrows.}
    \label{fig:ss_vs_cd}
\end{figure}

\begin{figure}
    \centering
    \includegraphics[width=\columnwidth]{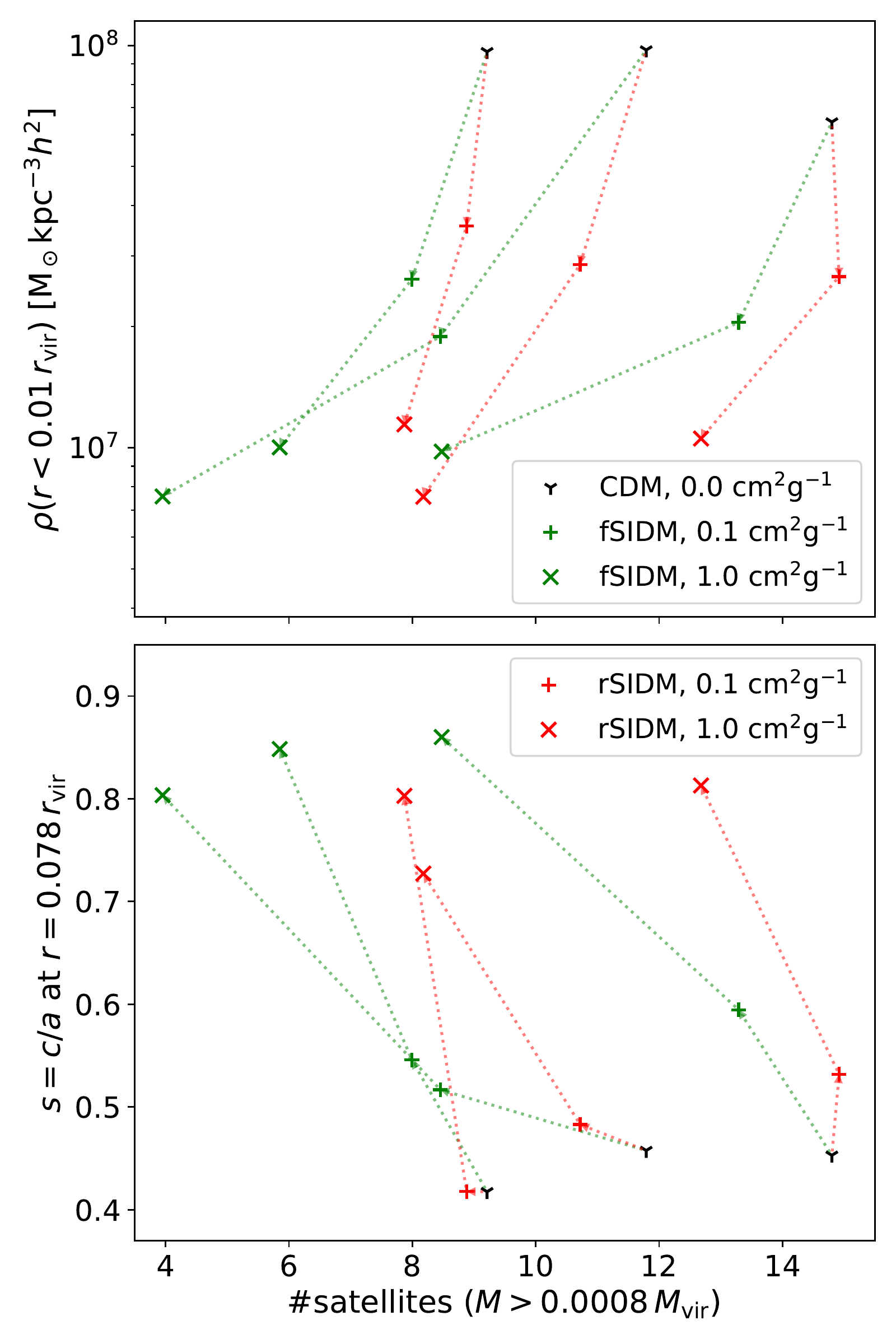}
    \caption{We show the three most massive haloes of our full-box simulation at $z=0$, the same as in Fig.~\ref{fig:ss_vs_cd}.
    The upper panel gives the central density within a radius of $0.01 \, r_\mathrm{vir}$ as a function of the number of satellites and the lower panel gives the shape of an ellipsoid that has the same volume as a sphere with a radius of $0.078 \, r_\mathrm{vir}$ as a function of the number of satellites.
    For the satellite count, only satellites within $r_\mathrm{vir}$ and with a mass larger than $0.0008 \, M_\mathrm{vir}$ were considered.
    This implies that every satellite is at least resolved by $\gtrsim 2200$ particles.
    The arrows connect the same systems across the simulations with various cross-sections.}
    \label{fig:fSIDMvsrSIDM}
\end{figure}

The main aim of this paper is to understand whether and how the phenomenology of rare and frequent self-interactions can help to distinguish between them.
In the following, we first investigate a potential degeneracy between rare and frequent self-interactions in terms of halo shape and central densities.
In Fig.~\ref{fig:ss_vs_cd}, we plot the shape as a function of central density for individual systems. The arrows indicate how they change when increasing the cross-section.
Here, we use the three most massive systems of the full-box simulation (uhr).
Although the values for the same object and cross-section differ from each other, this is not necessarily a qualitative difference.
Rather it can be interpreted as an issue of matching rare and frequent self-interactions.
As the two types of self-interaction roughly move in the same direction (in the shape--central density plane) when increasing the cross-section.

Secondly, we consider the number of satellites to find a qualitative distinction.
In Sec.~\ref{sec:halo_mass_func}, we found that it can make a difference for the abundance of satellites whether DM self-interactions are frequent or rare.
Compared to the differences, we found that for the central densities (Sec.~\ref{sec:density_profiles}) or the shapes (Sec.~\ref{sec:shapes}) the number of satellites seems to be more sensitive to the underlying form of the cross-section.
In Fig.~\ref{fig:fSIDMvsrSIDM}, we plot the central density (upper panel) and the shape (lower panel) as a function of the number of satellites for the same systems as in Fig.~\ref{fig:ss_vs_cd}.
We only count satellites that are within the virial radius and more massive than $0.0008 \, M_\mathrm{vir}$, which corresponds to a resolution of $\gtrsim 2200$ particles.
If fSIDM and rSIDM behave qualitatively the same, the systems should move along the same path in the plot when increasing the cross-section, regardless of whether the scattering is rare or frequent.
We use arrows to indicate how the systems change with increasing cross-section.
It becomes clear that fSIDM and rSIDM are not scaled versions of each other but are qualitatively different.
For a given number of satellites, we find the main halo to be less dense and more spherically symmetric for rSIDM compared to fSIDM.
If this qualitative difference is still present in full-physics simulations and strong enough to be observable, it could provide an avenue to distinguish between rSIDM and fSIDM.

A possible explanation of this difference between rSIDM and fSIDM goes as follows. 
In addition to tidal stripping (which is present even with collisionless DM), self-interactions between host and satellite particles can reduce the mass of, or even destroy, subhaloes as they orbit their host. This is because energy is transferred from the hot (high velocity dispersion) host halo into the cold (lower velocity dispersion) subhalo by self-interactions. In the case of rSIDM, this energy transfer takes the form of a small fraction of subhalo particles receiving large velocity kicks, which unbind them from the subhalo. In contrast, with fSIDM, all particles in the subhalo will interact with host halo particles, but will not receive kicks as large as in the rSIDM case. With the rSIDM and fSIDM cross-sections matched such that the same total energy transfer from host halo to subhalo takes place, and if we consider a time where the total transferred energy is just enough to unbind the subhalo, the fSIDM halo will spread this energy amongst all its particles and become unbound, while the rSIDM halo will lose some of this energy to subhalo particles that leave the subhalo at high velocity. In consequence, fSIDM satellites dissolve faster than their rSIDM equivalents.

\subsection{Constraints on fSIDM}
In this paper, we do not aim to compare our simulations to observations but previous studies of rSIDM may allow placing constraints on the velocity-independent fSIDM cross-section.

So far we have found that fSIDM and rSIDM behave similarly in many aspects.
In particular, if we consider the density (Fig.~\ref{fig:halo_density_profiles}) or shape (Fig.~\ref{fig:halo_shape_s}) profiles, we find that the momentum-transfer cross-section can roughly match fSIDM and rSIDM.
This can allow transferring constraints of the cross-section for rSIDM to fSIDM. In many situations, we found that fSIDM has a slightly larger effect than rSIDM, when the same momentum-transfer cross-section is considered. This implies that upper limits on the momentum-transfer cross-section may be even more stringent for the case of fSIDM. Given this, the established upper limits on rSIDM could be viewed as conservative limits for the case of fSIDM. In particular, this applies to our shape measurements for the whole mass range we have studied and to the density profiles for the two most massive mass bins ($\gtrsim 10^{13} \, \mathrm{M_\odot}$, see also Fig.~\ref{fig:central_density}).

In this context, the work of \cite{Sagunski_2021} is relevant. They studied the core sizes of galaxy groups and clusters to derive limits on the self-interaction cross-section. Their upper limits at a confidence level of 95\% on the total cross-section are $\sigma / m_\chi = 1.1 \, \mathrm{cm}^2 \, \mathrm{g}^{-1}$ for groups and $\sigma / m_\chi = 0.35 \, \mathrm{cm}^2 \, \mathrm{g}^{-1}$ for clusters. In terms of the momentum-transfer cross-section, this corresponds to $\sigma_{\tilde{T}} / m_\chi = 0.55 \, \mathrm{cm}^2 \, \mathrm{g}^{-1}$ for groups and $\sigma_{\tilde{T}} / m_\chi = 0.175 \, \mathrm{cm}^2 \, \mathrm{g}^{-1}$ for clusters.
As discussed above, these limits should hold for fSIDM too.

Another relevant study is \cite{Peter_2013}. They examined the shapes of galaxy clusters and consider a velocity-independent cross-section of $\sigma/m_\chi = 1.0 \, \mathrm{cm}^2 \, \mathrm{g}^{-1}$ (total cross-section) to be unlikely large. This corresponds to $\sigma_{\tilde{T}} / m_\chi = 0.5 \, \mathrm{cm}^2 \, \mathrm{g}^{-1}$ and should apply to fSIDM too. However, we have to note that \cite{Peter_2013} used DM-only simulations. \cite{Robertson_2019} showed that the shape of the intracluster medium (ICM) is hardly affected by the DM physics, but that the DM component becomes considerably rounder due to self-interactions, even if baryons are present. Furthermore, the presence of baryons makes the DM shapes become more round too, which rules in favour of excluding a cross-section of $\sigma_{\tilde{T}} / m_\chi \geq 0.5 \, \mathrm{cm}^2 \, \mathrm{g}^{-1}$ for rSIDM and fSIDM.

\section{Discussion} \label{sec:disscussion}

In this section, we discuss the limitations and implications of our results. 
Our findings depend partially on \textsc{subfind}, as we use the halo and subhalo positions as well as their mass and spatial extension.
Only a few of our results are independent of \textsc{subfind}, and these are the matter power spectrum and the density PDF.
We want to point out that there are a number of codes for identifying substructure \citep[e.g.][]{Knollmann_2009, Maciejewski_2009, Tweet_2009, Behroozi_2012, Han_2017, Elahi_2019}.
They employ different algorithms, which may give somewhat different results \citep{Knebe_2013}.

We found that the abundance of satellites is sensitive to differences between rare and frequent DM scatterings.
In particular, for low-mass satellites, the difference between the DM models becomes larger.
As such it would be interesting to study those low-mass satellites with a higher resolution.
In this context, a hybrid approach as presented in \cite{Zeng_2021} is an efficient technique that could help to improve our results.

In general, we would expect to find the largest differences between rSIDM and fSIDM in systems that are far from equilibrium.
However, there are systems that may be similar or even more sensitive to the shape of the differential cross-section. Mergers have received attention in the context of SIDM \citep{Randall_2008, Kahlhoefer_2014, Harvey_2015, Kim_2017b, Robertson_2017a, Robertson_2017b, Fischer_2021a, Fischer_2021b}.
It has been shown that the DM--galaxy offsets can produce discernible differences between DM models.
In addition to offsets, the distribution of a collisionless component (such as galaxies or stars) can give a handle to constrain DM properties.

The main simplification of our simulations is the neglect of baryonic matter and the associated physics. While DM-only simulations may provide reasonably accurate results on large scales, they fail on scales where galaxies form.
In particular, the inner regions of haloes can be strongly affected by baryons.
It is well known that feedback mechanisms from supernovae can create DM density cores too \citep{Read_2005, Governato_2012,Pontzen_2012,di_Cintio_2013, Brooks_2014, di_Cintio_2014, Pontzen_2014, Onorbe_2015, Tollet_2016, Benitez-Llambay_2019}, but also black holes \citep[e.g.][]{Martizzi_2013, Peirani_2017, Silk_2017}.
Recently, \cite{Burger_2021} studied degeneracies between cores produced by supernova feedback and SIDM cores. They found that the velocity dispersion profile produced by self-interactions is closer to isothermal profiles than in cores that are generated by supernova feedback in systems with bursty star formation.
Nevertheless, there are properties that are less affected by baryons.
We found that at large radii the halo shapes are more affected by self-interactions than the density profiles (see Appendix~\ref{sec:dslr}).
At these radii, baryons play a less important role than in the core region. 
\cite{Vargya_2021} studied an MW-like galaxy and proposed to compare the shape of the stellar and gas component to the shape of the total matter distribution at a radial range of 2--20 kpc.
Thus, the shapes at large radii (but still small enough to be affected by self-interactions) are more sensitive to DM physics than the central density or cores size because they are less influenced by baryonic physics.
The same could be true for the abundance of satellites, at least as far as DM-rich satellites are concerned.

Given a specific DM model, a single property can help constrain the momentum-transfer cross-section, but is quite limited in providing bounds on the angular dependence of the differential cross-section.
An observation, e.g.\ of the abundance of satellites, could be explained by rSIDM as well as by fSIDM with a different momentum-transfer cross-section.
In combination with another property, such as the shape, it can become possible to derive bounds on the typical scattering angle as demonstrated in Section~\ref{sec:fSIDMvsrSIDM}.
For a given DM model, multiple measurements could lead to bounds on $\sigma_\mathrm{\Tilde{T}}/m_\chi$, which are not compatible with each other and thus exclude the considered model.

In this context, we want to stress that matching cross-sections of various DM models is difficult because they typically behave qualitatively different.
Strictly speaking, it is impossible for rare and frequent self-interactions and probably for further model variations too. 

Several studies \citep{Kaplinghat_2016, Correa_2021, Gilman_2021, Sagunski_2021} have suggested that the self-interaction cross-section should be velocity-dependent, i.e.\ decrease for higher velocities.
We have not studied this, but a velocity dependence should lead to qualitatively different results and is a reasonable step to extend our study.
Moreover, velocity-dependent scattering is well motivated from a particle physics perspective.
This is, in particular, the case for light mediator models, which interact frequently \citep[e.g.][]{Buckley_2010, Loeb_2011, Bringmann:2016din}.

Another important step forward would be to include baryonic matter and its feedback processes.
At small radii, the mass distribution of haloes is typically dominated by the baryonic component.
As explained above, feedback mechanisms such as outflows from supernovae can alter the DM distribution and also produce cored profiles. By this, they can mitigate the core-cusp problem \citep[e.g.][]{Brooks_2013, Brooks_2014, Chan_2015, Onorbe_2015, ElBadry_2016} and the diversity problem \citep[e.g.][]{ElBadry_2018}.
Thus, it is important to take baryonic effects into account when constraining the properties of DM models using these small-scale problems.

\section{Conclusions} \label{sec:conlusion}

In this paper, we have presented the first cosmological simulations of DM with frequent self-interactions.
We have compared DM-only simulations of CDM, rSIDM, and fSIDM in terms of various measures, such as the density PDF, the matter power spectrum, the two-point correlation function of haloes and subhaloes, and the halo and subhalo mass functions.
In addition, we have investigated the density and circular velocity profiles of the DM haloes as well as their shapes.
Finally, we have examined qualitative differences between fSIDM and rSIDM.
Our main results from these simulations are as follows:

\begin{itemize}
    \item On large scales, rSIDM and fSIDM are very similar to CDM, but deviate on small scales.
    \item Regarding the suppression of small-scale structures, rSIDM and fSIDM behave very similarly.
    This includes the power spectrum, the density PDF, and the two-point correlation function as well as the density core formation and shapes of haloes.
    \item We found an interestingly large suppression of the abundance of satellites in fSIDM compared to rSIDM.
    \item It may be possible to distinguish observationally between rSIDM and fSIDM using a combination of measurements. One promising avenue is the combination of shape or density profile measurements with the abundance of satellites. Further investigations, such as full-physics simulations, are needed to find out whether observations can discriminate between these DM models.
    \item Rare and frequent self-interactions behave similarly in many aspects. This often allows transferring upper limits on the cross-sections of rSIDM to fSIDM.
\end{itemize}
We have conducted cosmological DM-only simulations to understand phenomenological differences between large- and small-angle DM scattering.
Our results may prove helpful for more sophisticated studies that compare simulations to observations with the aim to discriminate between rSIDM and fSIDM.
Such studies will include baryonic matter and baryonic physics, such as gas cooling, star formation, active galactic nuclei, and associated feedback mechanisms. This is the subject of forthcoming work.

\section*{Acknowledgements}
MSF thanks Lucas Valenzuela for the discussion on measuring halo shapes.
This work is funded by the Deutsche Forschungsgemeinschaft (DFG, German Research Foundation) under Germany's Excellence Strategy -- EXC 2121 ``Quantum Universe'' --  390833306, Germany’s Excellence Strategy -- EXC-2094 ``Origins'' -- 390783311, and the Emmy Noether Grant No.\ KA 4662/1-1. ARa acknowledges support from the grant PRIN-MIUR 2017 WSCC32.
KD acknowledges support by the COMPLEX project from the European Research Council (ERC) under the European Union’s Horizon 2020 research and innovation programme grant agreement ERC-2019-AdG 882679.
The simulations have been carried out on the computing facilities of the Computational Center for Particle and Astrophysics (C2PAP). Preprint numbers: TTP22-026 and DESY-22-072.

\textit{Software:}
\textsc{NumPy} \citep{NumPy},
\textsc{Matplotlib} \citep{Matplotlib},
\textsc{SciPy} \citep{SciPy}

\section*{Data Availability}

The data underlying this paper will be shared on reasonable request to the corresponding author.



\bibliographystyle{mnras}
\bibliography{references} 



\appendix

\section{Comoving Integration Test} \label{sec:comoving_integration_test}
Here, we introduce a test problem for the comoving integration of frequent self-interactions and demonstrate that our implementation works.

Similar to the deceleration problem in Newtonian space presented in \cite{Fischer_2021a}, we construct a deceleration problem in an expanding space.
Therefore, we have a background density modelled by many particles that are at rest (vanishing canonical momentum).
A test particle of the same mass as the background particles has initially a non-zero velocity and is travelling through the background density.
Due to the self-interactions, the test particle is scattering many times, which leads to a deceleration.
For the test simulation, we only use the first step, the deceleration as described in section~2.1 of \cite{Fischer_2021a}, but not the second step, which re-adds the energy lost in the first one (described in section~2.2).
This test problem is conducted without any other physics, i.e.\ gravity is not present.

\begin{figure}
    \centering
    \includegraphics[width=\columnwidth]{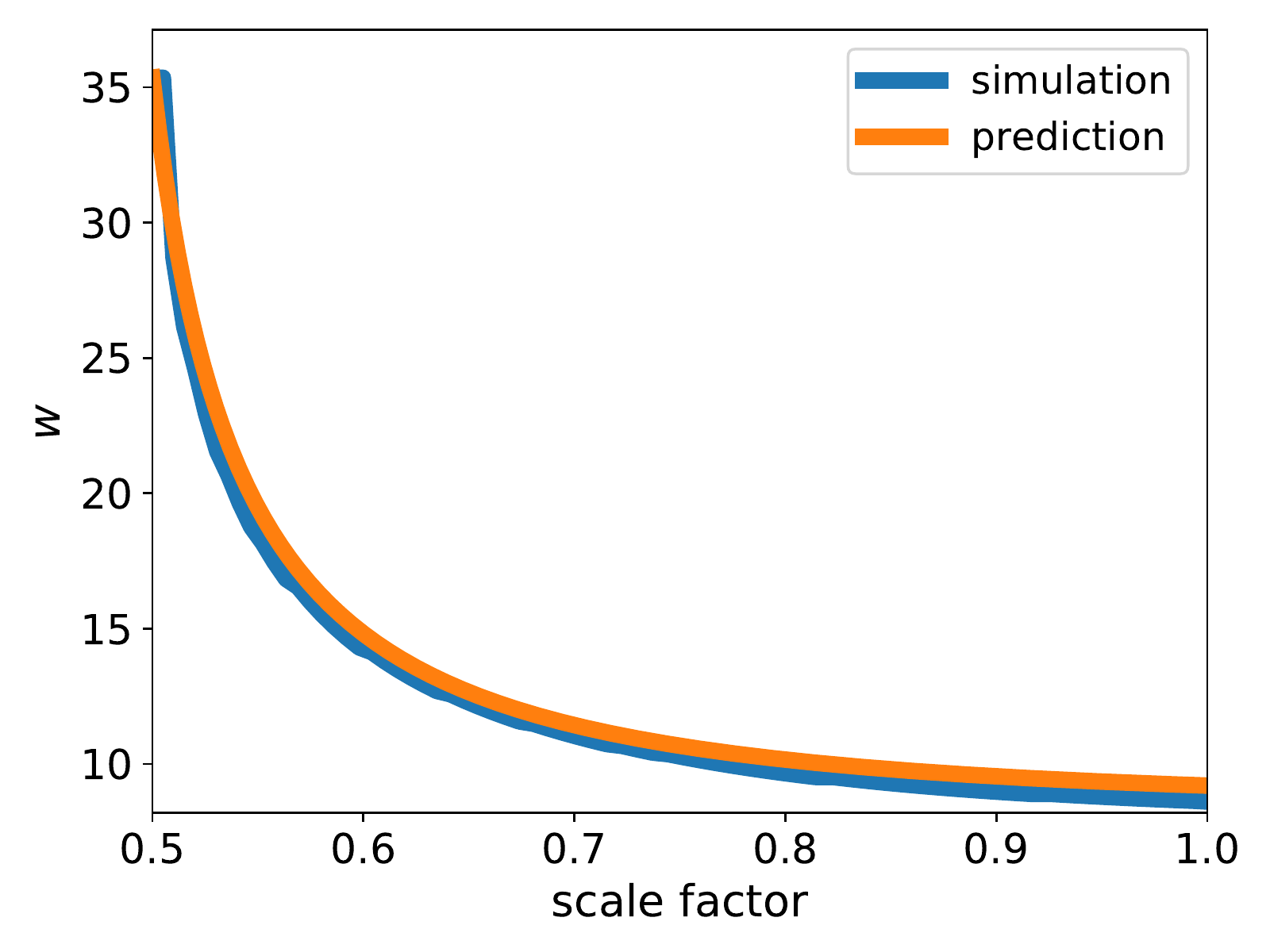}
    \caption{The cosmic deceleration problem in terms of the canonical momentum is shown.
    It is simulated from $a=0.5$ to $1.0$ with $122\,500$ particles in a cubic box with a comoving side length of $1400 \, \mathrm{kpc} \, h^{-1}$.
    The total mass is $22.8465 \times 10^{10} \, \mathrm{M_\odot} \, h^{-1}$, corresponding to a comoving density of $83.26 \, \mathrm{M_\odot}\,\mathrm{kpc}^{-3} \, h^2$.
    The initial snapshot velocity of the test particle is $100 \, \mathrm{kpc} \, \mathrm{Gyr}^{-1}$, which corresponds to an initial canonical momentum of $35.35534 \, \mathrm{kpc} \, \mathrm{Gyr}^{-1}$.
    The particles are evolved with a cross-section of $\sigma_\mathrm{\tilde{T}}/m_\chi = 5 \times 10^5 \, \mathrm{cm}^2 \, \mathrm{g}^{-1}$ and the SIDM kernel sizes are computed using $N_\mathrm{ngb} = 64$.}
    \label{fig:gadget_test_cosmo}
\end{figure}

In Fig.~\ref{fig:gadget_test_cosmo}, we show the cosmic deceleration problem by plotting the canonical momentum of the test particle as a function of the scale factor.
Note that in the absence of self-interactions the canonical momentum would stay constant over the cosmic expansion.
We can see that the simulation result matches the analytical prediction. Hence, we assume that the comoving integration is properly implemented.

\section{Convergence of density profiles} \label{sec:density_conv}

\begin{figure}
    \centering
    \includegraphics[width=\columnwidth]{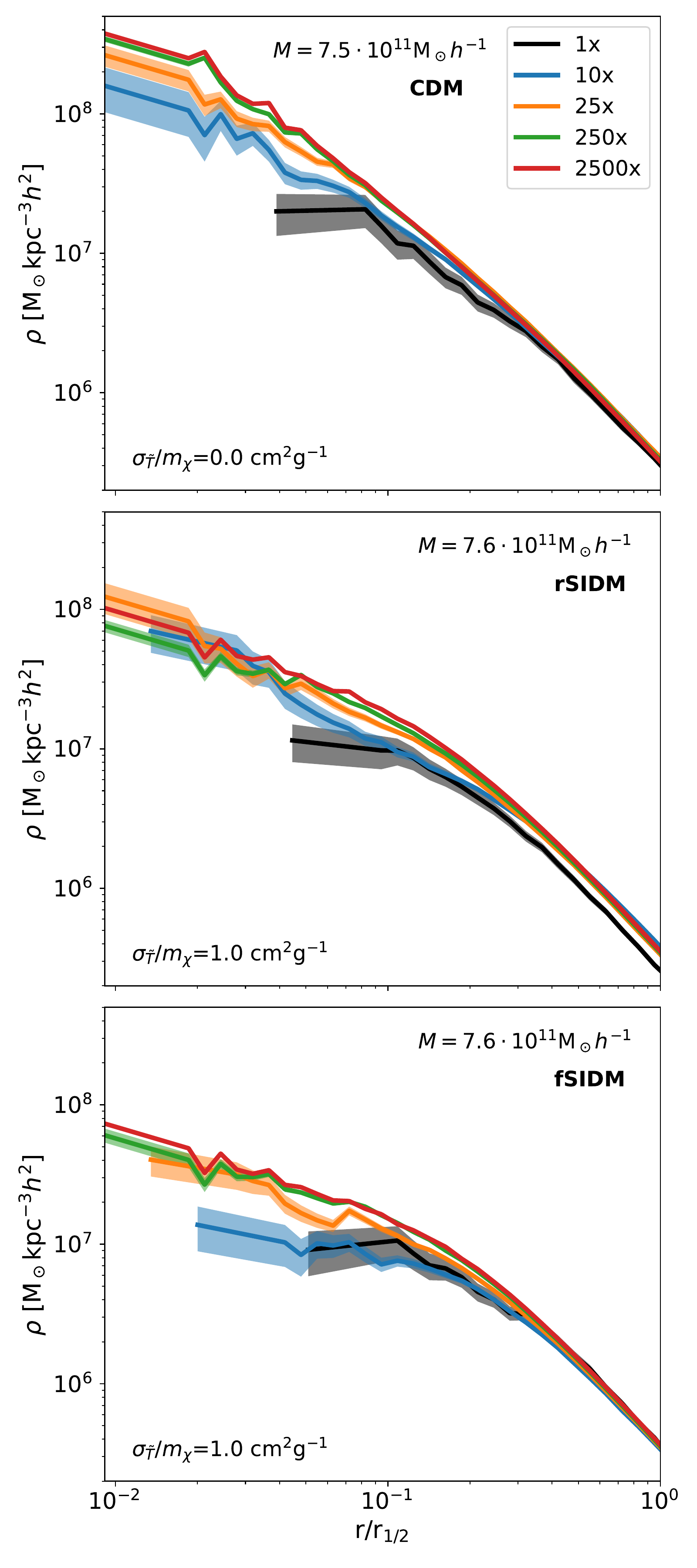}
    \caption{The density profile of the most massive subhalo in our zoom-in simulation is shown.
    In the top panel, we display the results for CDM, in the middle panel for rSIDM, and in the bottom panel for fSIDM.
    The colours indicate runs with different resolutions; further details can be found in Tab.~\ref{tab:sim_props_zoom}. In the highest resolution run, the halo is resolved by $\sim 2.3 \times 10^{6}$ particles.}
    \label{fig:density_profile_conv}
\end{figure}

In Fig.~\ref{fig:density_profile_conv}, we study the convergence of the density profile of the most massive subhalo in the zoom-in simulation.
We show the density for various resolutions and DM models as given in Tab.~\ref{tab:sim_props_zoom}.
For all three DM models, we find that the density profiles converge.
However, it seems that they converge at a different speed.
Comparing the two best resolved runs, CDM seems to converge the fastest, followed by fSIDM and rSIDM is the slowest.
The difference in the convergence speeds might be caused by the use of random numbers to model SIDM.
For fSIDM, they have a smaller influence on the particle trajectories than in rSIDM, which eventually could explain the deviation.

\section{Subhalo mass function} \label{sec:shmf2}

\begin{figure*}
    \centering
    \includegraphics[width=\columnwidth]{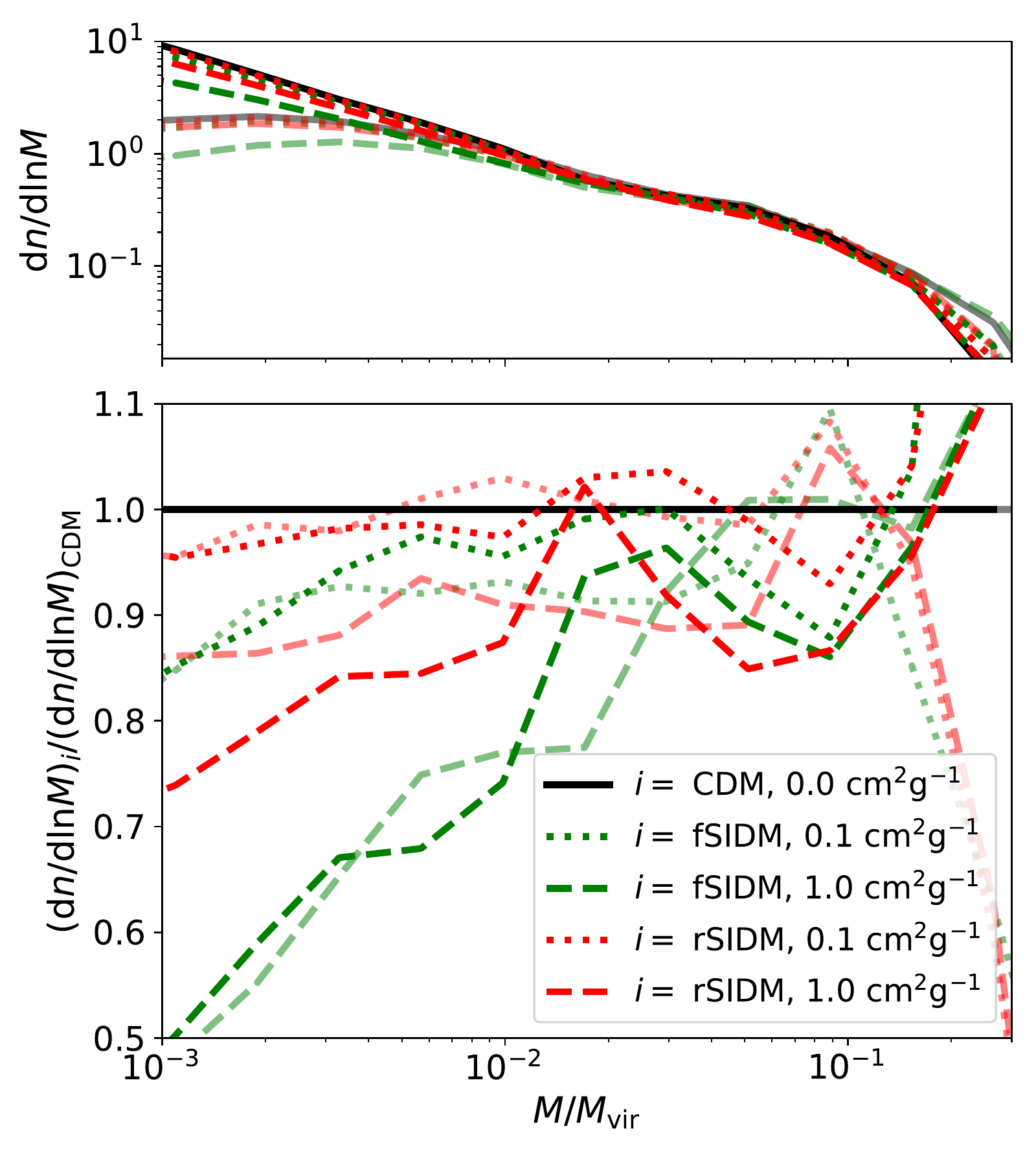}
    \includegraphics[width=\columnwidth]{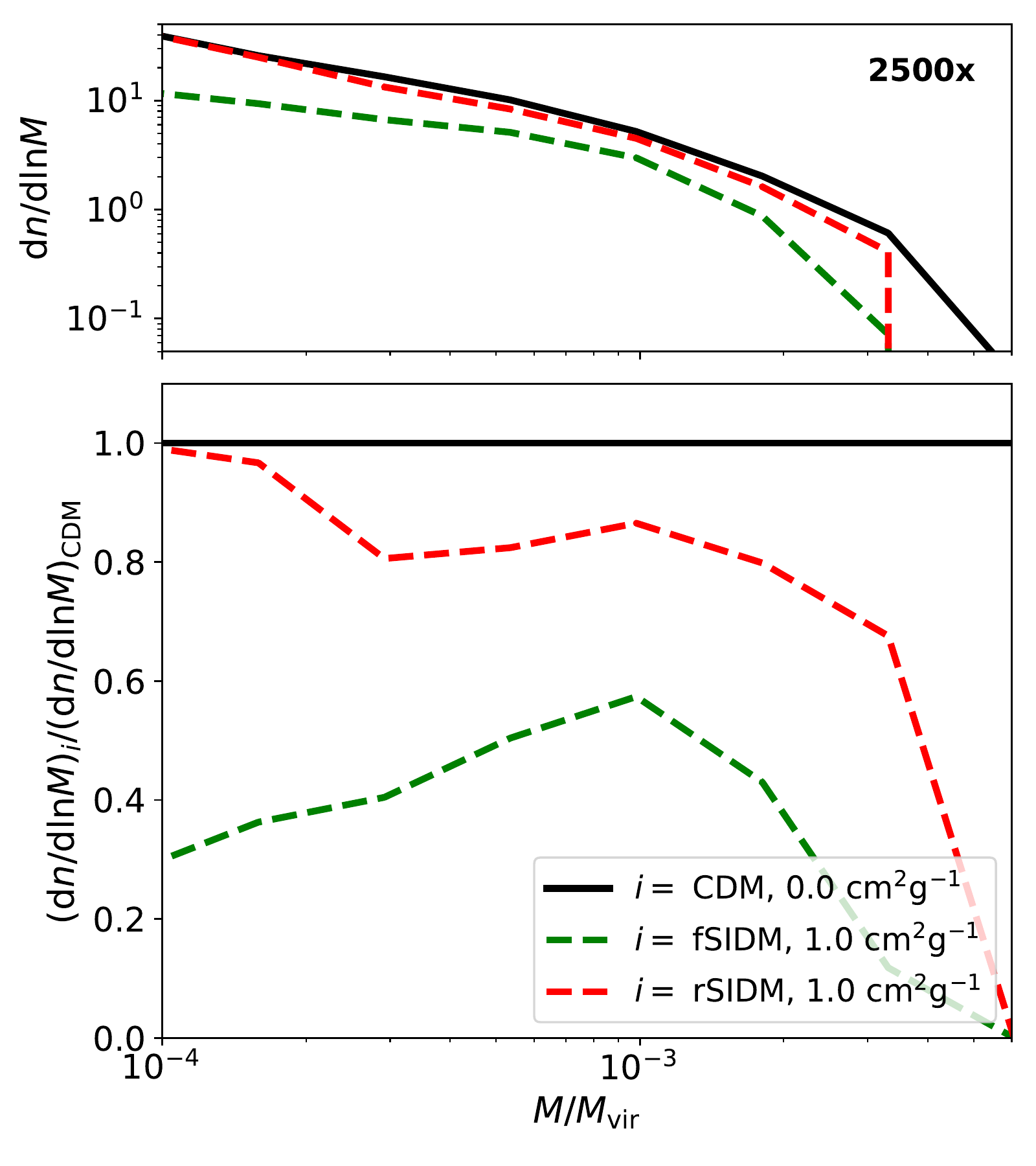}

    \caption{We display the number of satellites per logarithmic mass as a function of their total mass relative to the virial mass of their host.
    This is the same as in Fig.~\ref{fig:halo_shmf}, but with a selection radius of $1 \, r_\mathrm{vir}$.}
    \label{fig:halo_shmf2}
\end{figure*}

In Section~\ref{sec:halo_mass_func}, we studied the abundance of satellites as a function of mass. Here, we computed Fig.~\ref{fig:halo_shmf} with a smaller selection radius for the satellites. In Fig.~\ref{fig:halo_shmf2}, we only consider satellites within $1 \, r_\mathrm{vir}$ as we have done previously in Sec.~\ref{sec:fSIDMvsrSIDM}.

\section{Density and shape at larger radii} \label{sec:dslr}
\begin{figure}
    \centering
    \includegraphics[width=\columnwidth]{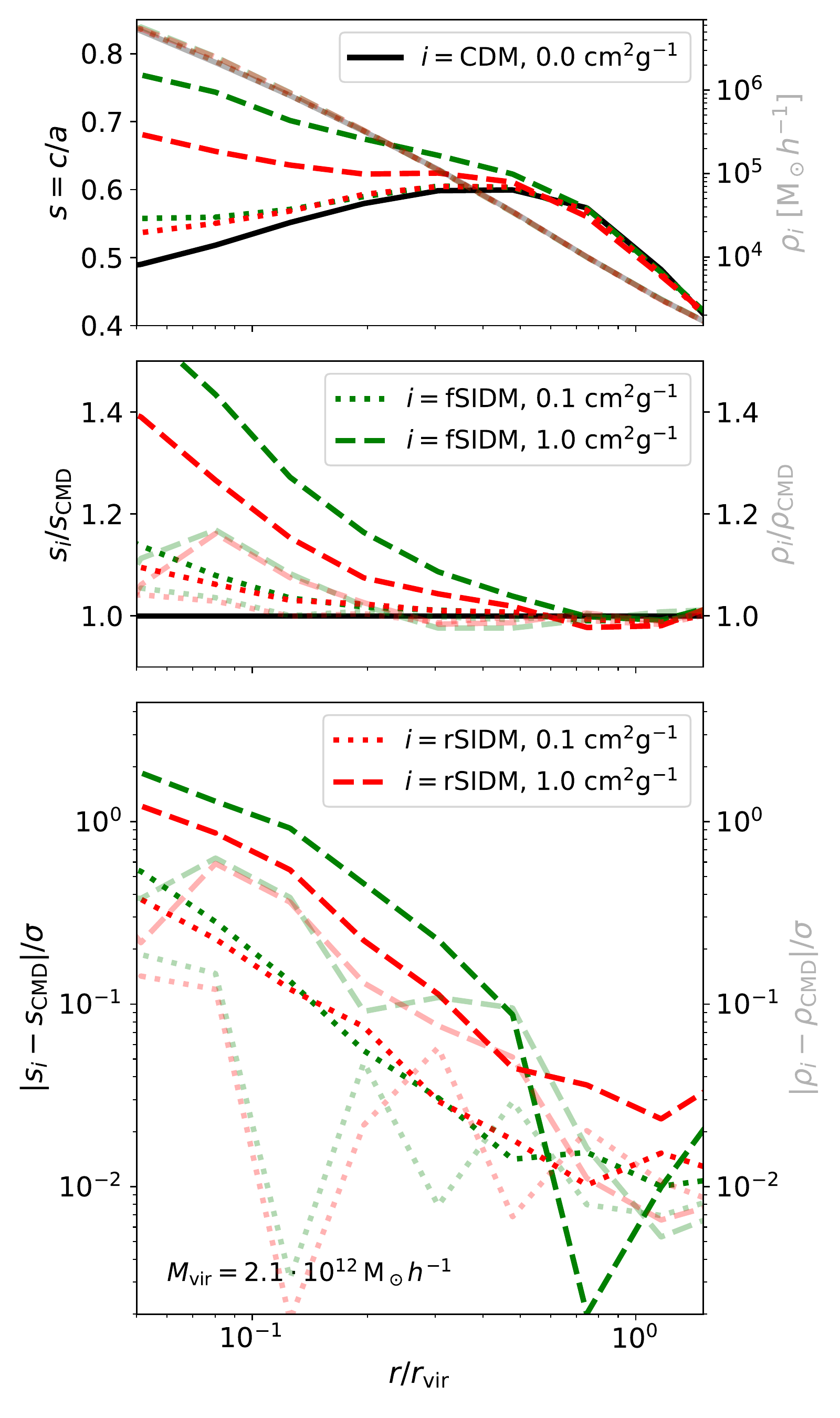}
    \caption{
    In the upper row, we display the median density (transparent) and shape profiles for the lowest halo mass bin (as in the bottom panels of Fig.~\ref{fig:halo_density_profiles}, \ref{fig:halo_shape_s}, and \ref{fig:halo_shape_t}) of our uhr simulations at $z=0$.
    The median mass of the haloes is $2.1 \times 10^{12} \, \mathrm{M_\odot}\, h^{-1}$.
    In the middle row, we show the ratio between the SIDM models and CDM.
    We further compute the difference of the SIDM models and CDM and divide it by the scatter among the individual systems. The result is displayed in the bottom row.
    The $x$-axis is in terms of the radius that a spherical shell with the same volume has.
    Here, we use the mean of the two shell boundaries.
    }
    \label{fig:dslr}
\end{figure}

In this appendix, we quantitatively evaluate how much the density and shape profiles differ at larger radii between the DM models.
We compute the shape of particles within elliptical shells using the tensor as given in Eq.~\eqref{eq:shape_tensor}.
As in Sec.~\ref{sec:shapes} we keep the volume of the shells during the iteration constant. Furthermore, we use the shells to compute the density profile too.
This is in contrast to the profiles shown in Fig.~\ref{fig:halo_density_profiles}, which were computed from spherical shells. 
For the computation, all particles are considered, not only those that belong to the halo as identified by \textsc{subfind}.
This implies that we also take the satellites into account, but it has almost no influence on the results shown here.
We use the scatter between the individual objects (16th and 84th percentile) to estimate how significant the deviation between the DM models is.
In Fig.~\ref{fig:dslr}, we show the results for the lowest mass bin of our uhr simulations.
We find that the difference between the models for radii beyond the density core is somewhat larger for the shapes than the density profiles (see the bottom row).
That seems to be true for $r\lesssim 0.25 \, r_\mathrm{vir}$ in the case of the smaller cross-section or $r\lesssim 0.4 \, r_\mathrm{vir}$ for the larger cross-section.
For most of the radial range we covered here, the SIDM shapes are rounder than the CDM shapes.
In contrast, the picture for the density profiles is less clear.
The density ratios are smaller (middle row) and the differences are noisier (bottom row).
For studying DM physics, this may make the measurement of shapes at larger radii preferable compared to densities.

\bsp	
\label{lastpage}
\end{document}